\documentclass[12pt]{article}

\usepackage[utf8]{inputenc}
\usepackage{cite}
\usepackage{color}
\usepackage{hyperref}
\usepackage[centertags]{amsmath}
\usepackage{amsfonts} 
\usepackage{amssymb}
\usepackage{amsthm}
\numberwithin{equation}{section} 
\usepackage{graphicx}
\definecolor{light-gray}{gray}{0.95}
\usepackage{subcaption}
\usepackage{abstract}
\begin{document}
\begin{titlepage}
\hfill \hbox{\footnotesize CERN-TH-2019-184},~
\hfill \hbox{\footnotesize NORDITA 2019-106},~
\hfill \hbox{\footnotesize QMUL-PH-19-33},~
\hfill \hbox{\footnotesize BOW-PH-168} 
\vskip 1.0cm
\begin{flushright}
\end{flushright}
\begin{center}
{\bf \Large 
A tale of two exponentiations   in \\[2mm]
${\cal N}=8$ supergravity at subleading level
}
\vskip 1.0cm

 \textsc{
Paolo Di Vecchia$^{a,b}$, 
Stephen G. Naculich$^c$, 
Rodolfo Russo$^d$, 
\\
Gabriele Veneziano$^{e,f}$, 
 and Chris D. White$^d$ 
}\\
 
\vspace*{0.8cm} $^a$ NORDITA, KTH Royal Institute of Technology and
Stockholm University, \\ 
Roslagstullsbacken 23, SE-10691 Stockholm, Sweden\\

$^b$ The Niels Bohr Institute, University of
Copenhagen, Blegdamsvej 17,\\
DK-2100 Copenhagen, Denmark\\

\vspace*{0.1cm} $^c$ Department of Physics, Bowdoin College, 
Brunswick, ME 04011 USA\\

\vspace*{0.1cm} $^d$ Centre for Research in String Theory, School of
Physics and Astronomy, \\
Queen Mary University of London, 327 Mile End
Road, London E1 4NS, UK\\

\vspace*{0.1cm} $^e$ Theory Department, CERN, CH-1211 Geneva 23, Switzerland\\

$^f$ Coll\`{e}ge de France, 11 place M. Berthelot,
75005 Paris, France\\

\end{center}

\vspace*{0.5cm}

\begin{abstract}
High-energy massless gravitational scattering in ${\cal N}=8$
supergravity was recently analyzed at leading level in the deflection
angle, uncovering an interesting connection between exponentiation of
infrared divergences in momentum space and the eikonal exponentiation
in impact parameter space. Here we extend that analysis to the first
non trivial sub-leading level in the deflection angle which, for
massless external particles, implies going to two loops, i.e. to third
post-Minkowskian (3PM) order.  As in the case of the leading eikonal,
we see that the factorisation of the momentum space amplitude into
the exponential of the one-loop result times a finite remainder hides
some basic simplicity of the impact parameter formulation. For the
conservative part of the process, the explicit outcome is infrared (IR)
finite, shows no logarithmic enhancement, and agrees with an old claim
in pure Einstein gravity, while the dissipative part is IR divergent
and should be regularized, as usual, by including soft gravitational
bremsstrahlung.  Finally, using recent three-loop results, we test the
expectation that eikonal formulation accounts for the exponentiation
of the lower-loop results in the momentum space amplitude. This passes
a number of highly non-trivial tests, but appears to fail for the
dissipative part of the process at all loop orders and sufficiently
subleading order in $\epsilon$, hinting at some lack of commutativity
of the relevant infrared limits for each exponentiation.
\end{abstract}

\vfill
\vskip 5.mm
\hrule width 5.cm
\vskip 2.mm
{
\noindent  {\scriptsize e-mails: divecchi@nbi.dk,naculich@bowdoin.edu,r.russo@qmul.ac.uk,gabriele.veneziano@cern.ch,christopher.white@qmul.ac.uk}}
\end{titlepage}

\section{Introduction }
\label{sec:intro}

The subject of gravitational collisions and radiation has been
receiving increased attention in recent years  particularly thanks to
the amazing experimental breakthroughs in gravitational-wave (GW)
detection \cite{Abbott:2016blz,Abbott:2016nmj,Abbott:2017vtc}.
From a theoretical standpoint one can tackle this problem
both at the classical General Relativity (CGR) level, through
numerical \cite{Lange:2017wki,Blackman:2017pcm} and analytical
\cite{Buonanno:1998gg,Buonanno:2000ef,Damour:2001tu,Damour:2000we}
methods, and at the quantum level using flat spacetime\footnote{An
exception is 't Hooft's 1987 calculation \cite{'tHooft:1987rb}
which  is carried out assuming a non-trivial background metric.}
calculations of scattering amplitudes. In this latter approach the
non-trivial classical spacetime geometry emerges from the resummation of
an infinite number of loop diagrams.  While the classical approach goes
back to the seventies \cite{DEath:1976bbo, Kovacs:1977uw,Kovacs:1978eu},
the quantum approach began in the late eighties with the above mentioned
work by 't Hooft \cite{'tHooft:1987rb} and independent parallel work by
two other groups \cite{Amati:1987wq,Muzinich:1987in,Amati:1987uf}
dealing with the transplanckian energy collisions of
strings in a generic number $D$ of macroscopic spacetime
dimensions. That approach was further developed in a number of
papers \cite{Amati:1990xe,Verlinde:1991iu,Amati:1992zb,Bellini:1992eb,
Amati:1993tb,Giddings:2010pp,Kabat:1992tb,Collado:2018isu,KoemansCollado:2019lnh}
and extended to the scattering of strings off a stack of D-branes
\cite{DAppollonio:2010krb,DAppollonio:2013mgj}. Many features of CGR,
such as deflection angles, time delays and tidal excitations, were neatly
recovered and new effects related to the finite string size were uncovered
\cite{Amati:1988tn,DAppollonio:2015fly}(see \cite{Veneziano:2015kfb}
for a recent review). In even more recent studies the method was
extended to the calculation of the gravitational bremsstrahlung
\cite{Gruzinov:2014moa, Ciafaloni:2015xsr, Laddha:2018vbn, Sahoo:2018lxl,
Ciafaloni:2018uwe, Addazi:2019mjh} produced in these ``gedanken
collisions". Other groups have used gauge theory and amplitude methods to examine
similar issues~\cite{Melville:2013qca,Luna:2016idw,Akhoury:2013yua,Bjerrum-Bohr:2016hpa,Bjerrum-Bohr:2018xdl,Bjerrum-Bohr:2019kec,Kalin:2019rwq}.

Although a priori the problem of transplanckian-energy collisions of light
particles or strings appears to be unrelated to the one of two coalescing
black holes, it has been stressed by Damour \cite{Damour:2017zjx} that
understanding such idealized processes can bring valuable information
about the parameters that enter the Effective-One-Body (EOB) potential
\cite{Buonanno:1998gg,Buonanno:2000ef,Damour:2001tu,Damour:2000we}
needed for the computation of the waveforms produced in actual
black-hole mergers\footnote{Recently, impressive amplitude
calculations have also been carried out for the collision of massive
(and typically non-relativistic) particles up the two-loop (3PM) order
\cite{Bern:2019nnu, Bern:2019crd} and their
outcome was incorporated into the EOB potential \cite{Antonelli:2019ytb}.}.

Irrespectively of their potential usefulness in GW research the problem
of high-energy gravitational scattering and radiation also presents
considerable theoretical interest. Indeed the original motivations for
such a study were quite disconnected from GW physics but rather related
to the problem of constructing a unitary gravitational S-matrix and thus
an explicit solution to the information puzzle in quantum black-hole
physics. So far that program has been only partly successful. It was
possible to show how, in the region of large impact parameters (small
deflection angles), the violation of tree-level unitarity is cured by loop
corrections even in the presence of string-size effects; at the opposite
end only a few interesting insights (see e.g. \cite{Amati:2007ak}) have
been achieved in the regime of small impact parameter (where gravitational
collapse is expected to occur) and the precise way unitarity is preserved
(if it is) is still somewhat mysterious \cite{Ciafaloni:2011de}.

The idea of this work is to start investigating such questions in the
context of a more manageable theory, ${\cal N}=8$ supergravity, which,
despite being different from CGR, should share with it the most
important large-distance (infrared) features. Hopefully, in this
highly supersymmetric context, one will be able to enter even the
gravitational collapse regime: after all the famous microscopic
understanding of black-hole entropy in string theory
\cite{Strominger:1996sh} does make crucial use of supersymmetry!

In a recent paper \cite{DiVecchia:2019myk} we have shown that the
exponentiation in impact-parameter space of the leading high-energy
($s \rightarrow \infty$) terms into a leading eikonal phase has non
trivial implications for the correction terms (the so-called remainders)
to another exponentiation, this time in momentum space, of infrared
divergences.  And indeed the two- and three-loop remainders of
\cite{Henn:2019rgj} are found to be fully consistent with those
implications. Here we extend that analysis to the first subleading
correction in the high-energy expansion of the eikonal phase
(equivalently a small-deflection-angle expansion). More precisely, we
focus on the scattering of transplanckian-energy {\it massless}
particles and check the validity of an extension of the leading
eikonal to include additional subleading contributions which can be
determined from the already known higher-loop amplitudes in ${\cal
  N}=8$ supergravity. For external massless states the even-order
loops provide new classical contributions to the eikonal
phase. Because of unitarity, they must exponentiate and therefore have
to be added to the leading eikonal phase obtained from tree
diagrams. By contrast, the odd-order loops provide only quantum
contributions and do not need to exponentiate; they must nonetheless be
included in the analysis because they mix at higher orders with the
classical contributions to reproduce the full scattering amplitude.

Such a procedure allows for a non-trivial consistency check by using again
the  three-loop  results of \cite{Henn:2019rgj} where we do not expect any
new classical contribution. Therefore, all the scattering data up to the
first subleading level in the high energy expansion should be reproduced
from the eikonal expansion. We find that the check works for all terms
except for a mismatch in the non logarithmically enhanced imaginary part
of the amplitude at order ${\cal{O}} (\epsilon^0 )$. More checks
of $b$-space exponentiation can be performed at all loops for the two
leading-$\epsilon$ terms. Although new mismatches are found to occur
we notice that they can all be absorbed in a relatively simple, but IR
singular, redefinition of the three-loop remainder. Possible origins of
these mismatches are discussed.

On the way we will also compute the first classical correction to the
eikonal phase (deflection angle) which, for massless-particle collision
only occurs \cite{Amati:1987uf} at the two-loop (or $3PM$) order and
compare it successfully with the one obtained long ago in pure Einstein
gravity \cite{Amati:1990xe}. The presence of a non-trivial classical
correction to the massless 3PM eikonal in ${\cal{N}}=8$ supergravity
represents a new result. This property is likely to persist also when
masses for external particles are introduced even in a supersymmetry
preserving way, as done in \cite{Caron-Huot:2018ape}. In the latter
work, it was shown that the 2PM eikonal vanishes in a maximally
supersymmetric setup also in the massive case. Moreover, it is possible
to perform a probe analysis if one of the masses is much bigger than
any other scale in the problem, for instance by using D6-branes as
done  in~\cite{DAppollonio:2010krb}: the result for the deflection angle
$\Theta_6$ in Eq.~(4.5) of~\cite{DAppollonio:2010krb} is consistent with
the assumption that all classical corrections to the leading eikonal
vanish in the probe limit for ${\cal N}=8$ supergravity. In view of
this result, one might have conjectured that the leading eikonal phase
(deflection angle) is exact for this theory even when both particles are
dynamical; the presence of a non-trivial correction at two-loop order
shows that this is not the case.

We also compute the non-conservative part of the subleading eikonal
(the leading being exactly conservative) which should be relevant for
understanding the accompanying gravitational radiation directly at the
quantum level. Actually, in the soft-graviton limit this should match
a calculation already carried out in \cite{Addazi:2019mjh} for ${\cal
N}=8$ supergravity.

The outline of the paper is as follows. In Sect. \ref{sec:leaddel} we
discuss the two types of exponentiations and the distinction between
classical and quantum contributions at arbitrary loop order. In
Sect. \ref{sec:leadexp} we summarize, for completeness, the check
presented in \cite{DiVecchia:2019myk} that the scattering data up to
three loops are consistent with the leading eikonal exponentiation. In
Sec. \ref{sec:subleaddel} we  extend the procedure to subleading terms at
high energy and then concentrate on the first subleading correction. Here
we find interesting results on the classical corrections at two-loop order
and compare them with those obtained in pure Einstein gravity. In
Sect. \ref{sec:allorders} we compare the two exponentiations at different
orders in $\epsilon$ and in the loop expansion. In particular, in
Sect. \ref{sec:elm1} we present successful checks for the first two terms
in the $\epsilon$ expansion (for which one can neglect the remainder
functions), while in Sect. \ref{sec:elm3} we consider the third and
fourth terms in the $\epsilon$ expansion, which are sensitive to the
two and three-loop remainders, and show that  a simple (but IR singular)
modification of the three-loop remainder cures all the mismatches.
Unlike in the previous
sections, in  Sect. \ref{sec:deld4} we perform  the calculation of  the
subleading eikonal phase directly in four dimensions. We find agreement
for the real phase at order ${\cal{O}} (\epsilon^0)$ (calculated from
arbitrary $D = 4 - 2 \epsilon$) and discuss the origin of a disagreement
on the imaginary part. In Sect. \ref{sec:concl} we summarize our results
and discuss some possible interpretation of the mismatches we found
between the two exponentiations. 
In Appendix \ref{AppB} we give some useful
formulas for the Fourier transforms used in the text and, in Appendix
\ref{AppC}, we write down for convenience the scattering data at two
and three loops extracted from Ref.\cite{Henn:2019rgj}.

\section{Two different kinds of exponentiation}
\label{sec:leaddel}

Amplitudes in ${\cal{N}}=8$ supergravity in four spacetime dimensions
continue to be at the centre of  intense investigation as they provide
the ideal laboratory to test ideas and techniques that then can be used
also in other, more physical, theories. Over the last few years the UV
properties of the ${\cal N}=8$ four-point amplitudes have been studied
to high-loop order, see for instance~\cite{Bern:2018jmv} and references
therein. In this paper we will focus on a complementary aspect of the
same scattering process: the high energy, small angle (Regge) regime. In
terms of the Mandelstam variables\footnote{Eq.~\eqref{eq:manv} assumes
that all external particles are incoming and the mostly plus metric,
but the remaining equations of this paper do not require  explicitly
this convention.}
\begin{equation}
  \label{eq:manv}
  s=-(k_1+k_2)^2\;,~~t=-(k_1+k_4)^2\;,~~u=-(k_1+k_3)^2 ~~;~~ s+t+u =0 \,,
\end{equation}
we work in the $s$-channel physical region ($s >0, t, u <0 $) and focus
on the near-forward regime $|t|\ll s$, hence we also have $|u|\gg |t|$.
In ${\cal N}=8$ supergravity the amplitude $A^{(\ell)}$ for four-particle
scattering at $\ell$ loops is proportional to the tree-level result. By
following the conventions of~\cite{BoucherVeronneau:2011qv,Henn:2019rgj}
we write the full amplitude as a formal series
\begin{equation}
  A (k_i, \ldots) = \sum_{\ell=0}^\infty A^{(\ell)}(k_i, \ldots) = A^{(0)} (k_i, \ldots) \left( 1+  \sum_{\ell=1}^\infty \alpha_G^{\ell} {\cal{A}}^{(\ell)} (t, s) \right)\;,
\label{LE1}
\end{equation}
where the dots stand for the dependence on the polarizations and
flavours of the external states, $A^{(0)}$ is the tree-level amplitude,
$A^{(\ell)}$ is the $\ell$-loop amplitude, ${\cal{A}}^{(\ell)}$ is its
``stripped" counterpart, and 
\begin{equation}
\alpha_G \equiv  \frac{G}{\pi \hbar} (4\pi\hbar^2)^\epsilon B(\epsilon) ~;
\quad \quad B(\epsilon)  \equiv \frac{  \Gamma^2 (1-\epsilon) \Gamma (1+\epsilon)}{\Gamma (1-2\epsilon)}\;,
\label{LE2}
\end{equation}
where $G$ is Newton's constant in $D=4-2\epsilon$ 
dimensions.\footnote{
Since the physical dimensions of $\alpha_G$ depend upon
$\epsilon$, specifically $[\alpha_G] \sim [{\rm energy}]^{-2+2\epsilon}$,
${\cal{A}}^{(\ell)}$
will have to exhibit the appropriate $\epsilon$-dependent dimensions
as well.
} 
 A simplification in  ${\cal{N}}=8$ supergravity is that the
loop expansion can be encoded in a set of ``scalar'' terms ({\rm i.e.}
the last factor in~\eqref{LE1}) that depend on $s$ and $t$, but not on
the other quantum numbers of the external particles.

We are interested in studying this dynamical factor and in understanding
whether there is an infinite subset of contributions that can be
expressed in a simple exponential form. A standard approach to find
an exponentiation is to use the infrared divergences as guidance: the
IR terms in the $\ell$-loop amplitude are entirely obtained from the
exponentiation in momentum space of the one-loop amplitude. Then it is
natural to rewrite~\eqref{LE1} in the  form
\begin{equation}
A (k_i, \ldots) = A^{(0)} (k_i,\ldots) \exp \Big( \alpha_G {\cal{A}}^{(1)} (t, s,\epsilon)\Big) \exp \left( \sum_{\ell=2}^{\infty} \alpha_G^{\ell} F^{(\ell)}(t, s, \epsilon) \right)\;,
\label{LE3}
\end{equation}
where we explicitly displayed the dependence on the dimensional
regularisation\footnote{Because of infrared divergences we have
performed all the calculations using dimensional regularization. This
procedure has been shown in Ref.~\cite{DiVecchia:2019myk} to be
essential to reproduce the high energy behavior of the scattering
amplitude.} parameter $\epsilon$ of the stripped one-loop amplitude
${\cal{A}}^{(1)}$ and the remainder function  $F^{(\ell)}$ whose study has been
initiated in~\cite{Naculich:2008ew,Brandhuber:2008tf}. As anticipated,
this formulation collects all infrared divergent contributions in the
exponential of ${\cal{A}}^{(1)}$, while all $F^{(\ell)}$ are expected
to be free from infrared divergences, {\rm i.e.} they are expected to
be finite as $\epsilon \to 0$.

A different approach is to look at the forward high-energy kinematics
({\rm i.e.} the Regge limit $|t|\ll s$). The leading contribution
to the $\ell$-loop amplitude $A^{(\ell)}$ scales as $s^{\ell
+2}$ with sub-leading contributions having, modulo logarithms,
lower powers of $s$ and higher powers of $t$. As mentioned in the
introduction, at sufficiently large $s$ such a perturbative behavior
violates partial wave unitarity ($Im a_J \ge |a_J|^2$, where $J$
is the angular momentum and $a_J$ is the $J^{\rm th}$ partial
wave amplitude~\cite{Soldate:1986mk,Giddings:2009gj}). Indeed,
the behaviour $A^{(\ell)}\sim s^{\ell +2}$ translates
into $a_J^{(\ell)} \sim  s^{\ell+1}$ which cannot satisfy
the above inequality at arbitrarily large $s$. It turns out
\cite{'tHooft:1987rb,Amati:1987wq,Muzinich:1987in,Amati:1987uf} that
unitarity is explicitly recovered at sufficiently large $J$ by means
of another kind of exponentiation, this time in impact parameter ($b
\sim 2 J/ \sqrt{s}$) --rather than  in transverse-momentum-- space as
in Eq.~(\ref{LE3}).

Let us start to see how this works in the case of the so-called leading eikonal approximation. 
It is convenient to extract from the tree-level amplitude the leading-energy behaviour
\begin{equation}
  A^{(0)}(k_i,\ldots) = A^{(0)}_L \hat{A}^{(0)}(k_i,\dots)\;,~~\mbox{with}~~~~
  A^{(0)}_L = \frac{8\pi \hbar  G s^2}{-t}\;.
\label{LE5}
\end{equation}
By construction, in the case of an elastic scattering, $\hat{A}^{(0)}$
starts with $1$ plus terms that are subleading in the $|t|\ll s$
limit. The leading behaviour $A^{(0)}_L$ in~\eqref{LE5} is the only
information we need about the tree-level amplitude.

In order to rewrite the leading energy results in impact parameter
space, we first introduce an auxiliary $(D-2)$-dimensional momentum
$q$ such that $q^2=|t|$. Then we take the Fourier transform where $b$
is the conjugate variable to $q$ and define the leading eikonal phase
by  \cite{Amati:1987wq,Muzinich:1987in,Amati:1987uf}:
\begin{equation}
2i \delta_0 (s, b)
= \int \frac{ d^{D-2} q}{ (2\pi\hbar)^{D-2}} 
{\rm e}^{ibq/\hbar} \frac{iA_L^{(0)}}{2s} 
= \,-\,{i G s \over \epsilon \hbar} \Gamma (1-\epsilon) (\pi b^2)^\epsilon\;,
\label{LE6}
\end{equation}
where we used Eq.~\eqref{B1}.  At one loop, we have
\begin{equation}
  A^{(1)} = A^{(0)} \alpha_G {\cal{A}}^{(1)} 
\longrightarrow 
A_L^{(0)} \alpha_G \left( \frac{-i \pi  s}{\epsilon (q^2)^{\epsilon}}\right) \equiv A_L^{(1)}\;,
\label{LE9}
\end{equation}
where in the step indicated by the arrow we focused on the leading term of~\eqref{SUM4} in the Regge (high energy) limit. By going to impact parameter space one gets:
\begin{equation}
\int \frac{d^{D-2} q}{(2\pi\hbar)^{D-2}} {\rm e}^{ibq/\hbar} \frac{i A_L^{(1)}}{2s} 
=
\int \frac{d^{D-2} q}{(2\pi\hbar)^{D-2}} {\rm e}^{ibq/\hbar} \frac{i A_L^{(0)}}{2s} \alpha_G \frac{-i \pi  s}{\epsilon (q^2)^{\epsilon}} 
= - \frac{1}{2} (2\delta_0)^2 \, .
\label{LE10}
\end{equation}
Thus we see that the sum of leading energy contributions of the tree
and one-loop amplitudes starts to exponentiate in impact parameter space
\begin{equation}
\int \frac{d^{D-2} q}{(2\pi\hbar)^{D-2}} {\rm e}^{ibq/\hbar} 
\left( 
\frac{i A_L^{(0)}}{2s} 
+ \frac{i A_L^{(1)}}{2s} 
+ \ldots 
\right)
= 2i\delta_0 - \frac{1}{2} (2\delta_0)^2+\ldots = {\rm e}^{2i\delta_0 (s, b)} -1 \;.
\label{LE11}
\end{equation}
Such an exponentiation works at all orders and resums all the terms of
order $(Gs)^{\ell}$. As a result we have recovered (elastic) unitarity
since we managed to lump all the divergent contributions at high energy
into a large phase:
\begin{equation}
 \label{LE}
  \frac{i A_L}{2s} = \int\! { d^{D-2} b} \, {\rm e}^{-ibq/\hbar} \left( {\rm e}^{2i\delta_0 (s, b)} -1 \right)\;.
\end{equation}
Note that this leading eikonal resummation should hold at any $D$
and is thus conceptually unrelated to the exponentiation of infrared
divergences. In Sect.~\ref{sec:leadexp} we will recall how such an
exponentiation agrees with explicit amplitude calculations up to three
loops.  In view of extending such an analysis to the first subleading
term in Sect.~\ref{sec:subleaddel} we anticipate here some general
considerations about exponentiation in impact-parameter space.

For this purpose it is convenient to associate with the centre of mass energy $\sqrt{s}$ a length scale:
\begin{equation}
R \equiv (G\sqrt{s})^{\frac{1}{1-2\epsilon}}\, , \qquad {\rm i.e.} \qquad ~ G\sqrt{s} \equiv R^{D-3} \, ,
\label{defR}
\end{equation}
in analogy with the Schwarzschild radius of CGR\footnote{ The actual Schwarzschild radius $R_S$ of a black hole of mass $\sqrt{s}$ differs from $R$ by a well-known $\epsilon$-dependent factor. Note that $R$ has the dimension of a length for any $\epsilon$.}. 
In the spirit of~\cite{Kosower:2018adc} we can now express the scaling of different terms at a given loop order in terms of the CGR quantities $b$ and $R$ and of Planck's constant.
The Fourier transform of the leading energy contribution to the $\ell$-loop amplitude scales as: 
\begin{equation}
\int \frac{d^{D-2} q}{(2\pi\hbar)^{D-2}} {\rm e}^{ibq/\hbar} 
\frac{i A_L^{(\ell)}}{2s} 
\sim \left[ \left( \frac{R}{b} \right)^{-2\epsilon} \frac{R \sqrt{s}}{\hbar} \right]^{\ell+1}
\label{scale}
\end{equation}
i.e. as the $(\ell+1)th$ power of the leading eikonal phase $\delta_0$ in \eqref{LE6}: 
\begin{equation}
 \delta_0 \sim  \frac{R \sqrt{s}}{\hbar} \left(\frac{R}{b}\right)^{-2\epsilon} \sim \frac{b \sqrt{s}}{\hbar} \left(\frac{R}{b}\right)^{1-2\epsilon}\, .
\label{scaling}
\end{equation}
This confirms that the leading eikonal resums arbitrarily high powers of
$\hbar^{-1}$ into an ${\cal O}(\hbar^{-1})$ phase provided we consider, in
order to make contact with CGR,  $R$ and $b$ as classical quantities. Of
particular relevance is the derivative of the eikonal phase with respect
to $b$ since it provides, via a saddle point estimate of the inverse
Fourier transform, the classical deflection angle to leading order in
$R/b$: $\theta_s \sim \left(\frac{R}{b}\right)^{1-2\epsilon}$. Such
a classical interpretation would fail if the resummation of all the
leading powers of  $\hbar^{-1}$ were not to exponentiate.  The last
term in~\eqref{scaling} is particularly suggestive since the quantity $b
\sqrt{s}$ can be identified, at the leading eikonal level, with the total
angular momentum of the process, assumed to be much larger than $\hbar$.

Let us now consider also the subleading energy contributions. The
amplitude  consists of a sum of  terms having powers of $s$ all the way
up to the leading power $\ell+1$. Each one of these terms behaves in
impact parameter space as follows (again neglecting possible logarithmic
enhancements):
\begin{eqnarray} 
\int \frac{d^{D-2} q}{(2\pi\hbar)^{D-2}} {\rm e}^{ibq/\hbar} 
\frac{iA^{(\ell)}}{2s} 
&\sim& \sum_{m=0} \hbar^{2m-\ell -1} G^{\ell+1} s^{\ell+1 -m} b^{2\epsilon (\ell+1)-2m} 
\nonumber \\
&=&
\sum_{m=0} \left( \frac{R}{b}\right)^{2m - 2\epsilon(\ell+1)} \left(\frac{R \sqrt{s}}{\hbar}\right)^{\ell+1 - 2m}\; .
\label{scalinggenl}
\end{eqnarray}
In the massless
case under consideration, and in $D=4$, the amplitude $A^{(\ell)}$
cannot depend on fractional powers of $s$. In particular, it does not
contain terms proportional to\footnote{We  take this as an empirical
fact whose deeper reason should rest on the fact that each power of $G$
must be accompanied by an (energy)$^2$ factor. In the absence of masses,
a non-integer power of $s$ would have to be accompanied by a non-integer
power of $t$ and/or $u$, producing a multiple discontinuity excluded
by Steinmann-relation-type arguments.} $\sqrt{s}$ and so the expansion
above is in terms of even powers $1/b^{2m}$, while in the massive
case all powers of $1/b$ can (and do) appear. In both
the massive and the massless cases, terms proportional to $1/\hbar$
must be themselves exponentiated through higher-loop contributions and
contribute to a classical correction to the eikonal $\delta$, while
contributions with higher powers of $1/\hbar$ must be accounted for by
the exponentiation of terms appearing at lower-loop order.

In particular, if $\ell$ is even,  the term with  $m = \frac{\ell}{2}$
is a new classical contribution to the eikonal, while the terms with $m <
\frac{\ell}{2}$ reconstruct the exponentiation of terms appearing at a
lower-loop order.  All other terms with non-negative powers of $\hbar$
are quantum terms and do not need to exponentiate.  If instead $\ell$
is odd, all terms with  $m \leq \frac{\ell-1}{2}$ contribute to the
exponentiation of terms appearing at lower loops, while the terms with
$m \geq \frac{\ell+1}{2}$ are quantum and do not necessarily exponentiate.

In conclusion, terms with $m<\frac{\ell}{2}$ do not contain new
information as far as the classical scattering is concerned and a
first ingredient relevant for the classical eikonal (and thus to the
deflection angle) appears in the massless case at each even-loop order
$A^{(2\ell)}$ for $m=\frac{\ell}{2}$. The odd-loop orders $A^{(2\ell+1)}$
do not contribute directly to the classical phase or angle. However they
still take part in the exponentiation and so are important to extract
the correct classical eikonal phase.

On the basis of these considerations we propose the following extension
of the leading eikonal to include also subleading contributions:
\begin{equation}
 \label{intro1}
\frac{i A (k_i,\ldots)}{2s} \simeq \hat{A}^{(0)}(k_i,\ldots) \int\! { d^{D-2} b} ~{\rm e}^{-ibq/\hbar} \, \left[\Big(1+2i\Delta(s,b) \Big)\, {\rm e}^{2i\delta (s, b)} -1\right] \;,
\end{equation}
where all the terms appearing in ${\rm e}^{2i\delta (s, b)}$ are
proportional\footnote{This resembles very much a WKB approximation
in which the ${\cal{O}}(\hbar^{-1})$ exponent contains a classical
action satisfying the  Hamilton-Jacobi equation. For a review of
the relationship between the WKB and eikonal approximations, see
e.g. ref.~\cite{weinberg_2015}.} to $\hbar^{-1}$ while those appearing
in the prefactor $\Delta$ contain the contributions with non-negative
powers of $\hbar$.
The use of $\simeq$ here and below indicates that the
identity~\eqref{intro1} is restricted to the non-analytic terms as $q\to
0$ that capture long-range effects in impact parameter space. Checking
the validity of~\eqref{intro1} will be one of the main themes of the
following sections.


\section{Check of (and constraints from) the leading-eikonal}
\label{sec:leadexp}

As argued in the previous section,  it is natural to assume that
the leading high energy contribution at any loop order is simply
captured  by taking the Fourier transform of the leading eikonal back
to momentum space, see~\eqref{LE}.  In ref.~\cite{DiVecchia:2019myk},
we showed that this equation reproduces the leading terms at two- and
three-loop order by using the full results for these amplitudes obtained
in refs.~\cite{BoucherVeronneau:2011qv,Henn:2019rgj}. This should hold at
any order in $\epsilon$ and not just for the contribution that survives
in $D=4$, and we provided evidence of this by checking~\eqref{LE} at
various orders in the $\epsilon$ expansion.

Let us now recall how the two exponentiations~\eqref{LE3} and~\eqref{LE}
are related. We focus on elastic processes where $\hat{A}^{(0)}$ is just
the identity operator ensuring that the in and the out states have the
same polarization and flavour; then, by starting from~\eqref{LE}, we have
\begin{equation}
\frac{i A_L}{2s} = \int\! { d^{D-2} b} \, {\rm e}^{-ibq/\hbar}
  \left(\sum_{\ell=1}^\infty \frac{1}{\ell!} (2i \delta_0 (s,b))^{\ell}\right)\;.
\label{LE13}
\end{equation}
The Fourier transform can be performed term by term thanks to~\eqref{B1bis} and by taking the tree-level result as an overall factor, we obtain
\begin{align}
  \label{LE14}
\frac{i A_L}{2s} = 
& \frac{i A_L^{(0)}}{2s} \sum_{\ell=0}^\infty \frac{1}{\ell!} 
\left[ - \frac{ i G s  }  {\epsilon\hbar}
 \Gamma ( 1- \epsilon)
\left( \frac{4\pi\hbar^2 }{q^2}\right)^{\epsilon} 
\right]^{\ell} 
\frac{ \Gamma ( \ell \epsilon +1 ) \Gamma (1 -\epsilon)}{\Gamma (1 -(\ell+1) \epsilon )} \\ \nonumber
  = & \frac{i A_L^{(0)}}{2s} \sum_{\ell=0}^\infty \frac{\alpha_G^\ell}{\ell!} \left( \frac{-i \pi s}{\epsilon (q^2)^\epsilon}\right)^\ell G^{(\ell)}(\epsilon)\;,
\end{align}
where
\begin{align}
G^{(\ell)}(\epsilon) &= \frac{\Gamma^\ell (1-2 \epsilon) \Gamma (1+\ell \epsilon)}{\Gamma^{\ell-1} (1 -\epsilon) \Gamma^\ell (1+ \epsilon) \Gamma (1- (\ell+1) \epsilon)}\;
\nonumber \\
& = 1 -\frac{1}{3} \ell \left(2 \ell^2+3 \ell-5\right) \zeta_3 \epsilon^3 
+ {\cal O} (\epsilon^4). 
 \label{eq:Gedef}
\end{align}
We can now compare this result with the exponentiation~\eqref{LE3} and in particular we focus on the 
two- and three-loop amplitudes that were studied in detail in~\cite{BoucherVeronneau:2011qv,Henn:2019rgj}
\begin{align}
  & \frac{1}{2}({\cal{A}}^{(1)}_L)^2 + F_L^{(2)}  = \frac{1}{2!}
  \left(\frac{-i \pi s}{\epsilon (q^2)^\epsilon}\right)^{2} G^{(2)} ,
  \label{LE12}  \\ 
& \frac{1}{3!}  ({\cal{A}}_L^{(1)})^3+ F_L^{(3)} + {\cal{A}}_L^{(1)} F_L^{(2)} =  \frac{1}{3!} \left(\frac{-i \pi s}{\epsilon (q^2)^\epsilon}\right)^{3} G^{(3)} ,
\label{LE12b}
\end{align}
where 
on the left-hand side we have the high energy expansion of~\eqref{LE3}
at two and three loops while on the right-hand side we have the
corresponding order as it appears in~\eqref{LE}.  Solving for $F_L^{(2)}
$ using ${\cal{A}}_L^{(1)} = \frac{-i \pi  s}{\epsilon (q^2)^{\epsilon}}$
from Eq.~\eqref{LE9},  we have
\begin{equation}
  F^{(2)}_L = \lim_{s \rightarrow \infty} F^{(2)} =  \frac{1}{2}  \left( \frac{-i \pi s}{\epsilon (q^2)^\epsilon}\right)^2  \left[G^{(2)} (\epsilon )-1\right]\;.
\label{LE17}
\end{equation}
Using Eq.~\eqref{eq:Gedef} we obtain
\begin{equation}
   F_L^{(2)} =  3 \pi^2 s^2 \epsilon \zeta_3 + {\cal{O}} (\epsilon^2, s) 
\label{LE20}
\end{equation}
in agreement with the first line of Eq.~(6.5) of
ref.~\cite{Henn:2019rgj}. Notice that, in the high energy expansion, the
contribution in~\eqref{LE20} is leading, {\rm i.e.} at the same level as 
$({\cal{A}}_L^{(1)})^2$, showing explicitly that the formulation
of~\eqref{LE3} does not collect all the leading energy terms in the
exponential factor.

At the next order in perturbation theory (three loops) the remainder
function contains a leading energy contribution also in the IR finite term.
From Eq.~\eqref{LE12b} we have
\begin{equation}
 F_L^{(3)} = \lim_{s \rightarrow \infty} F^{(3)} =  \frac{1}{3!} \left( \frac{-i \pi s}{\epsilon (q^2)^\epsilon} \right)^3 
\left[ (G^{(3)}-1) - 3 \left(G^{(2)}-1\right) \right]\;.
\label{LE22}
\end{equation}
Again using \eqref{eq:Gedef} one obtains 
\begin{eqnarray}
F_L^{(3)} = - \frac{2i}{3} \pi^3 s^3 \zeta_3 + {\cal{O}} (\epsilon, s^2)\;,
\label{LE25}
\end{eqnarray}
which agrees with the second line\footnote{In~\cite{Henn:2019rgj} $s$ is taken to be negative. In order to match the different conventions one can use the following dictionary for the Mandelstam variables $s_{HM}=u$, $u_{HM}=s$, $t_{HM}=t$, where the subscript HM indicates the variables used in~\cite{Henn:2019rgj}.} of Eq. (6.5) of ref.~\cite{Henn:2019rgj}.

\section{Exponentiation at the first  subleading eikonal}
\label{sec:subleaddel}

In this section we focus on the first subleading-energy correction to
the eikonal exponentiation.  

 As a first step, we need a better approximation
to ${\cal{A}}^{(1)}$, including  the subleading ${\cal O}(t/s)$
corrections.
It is possible to perform the massless one-loop box integral for general 
values of $\epsilon$ and of the kinematic variables, 
and then perform the Regge limit of the exact expression up to the desired order in $t/s$. 
A convenient starting point for such an expansion is ~\cite{Bern:1993kr},\cite{Bern:1998ug}:
\begin{eqnarray}
 \epsilon^2   {\cal{A}}^{(1)}  
= &&   (-s)^{-\epsilon}\left[ u ~F(\epsilon, 1+\frac{s}{t} ) + t~
F(\epsilon, 1+\frac{s}{u} )\right] \nonumber \\
&& + (-t)^{-\epsilon}\left[ u ~F(\epsilon, 1+\frac{t}{s} ) + s~
F(\epsilon,  1+\frac{t}{u} )\right] \nonumber \\
&& +(-u)^{-\epsilon}\left[ t ~F(\epsilon,  1+\frac{u}{s} ) + s~
F(\epsilon,  1+\frac{u}{t} )\right] \;,
\label{A1}
\end{eqnarray}
where $F(\epsilon, z) \equiv  {}_2 F_1 (1, -\epsilon; 1-\epsilon; z )$.
By following 
this approach and keeping track carefully of the phases due to the branch 
cuts of the amplitudes, one finds a closed and rather simple expression 
for ${\cal{A}}^{(1)}$: 
\begin{eqnarray}
{\cal{A}}^{(1)} &  = &  - \frac{i \pi s}{\epsilon (q^2)^\epsilon} + {\cal{A}}^{(1)}_{SL} ~+~\ldots  \;, 
\nonumber \\
{\cal{A}}_{SL}^{(1)} &  = &  
  \frac{q^2 (1+2 \epsilon)}{\epsilon (q^2)^\epsilon}  \left(
\log \frac{q^2}{s} + H(\epsilon) \right)  
~-~ \frac{2q^2 (2\epsilon +1)}{\epsilon^2 (\epsilon +1) s^\epsilon}  \cos^2  \frac{\pi \epsilon}{2} 
\nonumber \\
& &  ~+~ i \frac{\pi q^2}{\epsilon} \left[ \frac{1+\epsilon}{(q^2)^\epsilon} ~-~ \frac{1+2\epsilon}{s^\epsilon (1+\epsilon)} \frac{\sin \pi \epsilon}{\pi \epsilon} \right] \;,
\label{SUM4}
\end{eqnarray} 
where ${\cal A}_{SL}^{(1)}$ is the subleading level contribution in the eikonal limit 
of the stripped amplitude ${\cal A}^{(1)} $, the dots stand for terms of order $q^4 s^{-1}$, 
and we have defined
\begin{equation}
H(\epsilon) 
\equiv \psi (-\epsilon) - \psi (1) - 1 + \pi \cot \pi \epsilon \, ,
\label{SUM5}
\end{equation}
where $\psi(z)=\frac{d\ln(\Gamma(z))}{dz}$ is the Digamma function (for our purposes it is useful to recall that it satisfies $\psi(1)=-\gamma_E$, where $\gamma_E$ is the Euler–Mascheroni constant). Notice that the quantity defined in~\eqref{SUM5} diverges as  $\frac{2}{\epsilon}$ for small $\epsilon$.
This expression is valid for general 
values of $\epsilon$ up to the subleading level in the Regge limit and we 
checked that in this regime it reproduces the data of~\cite{Henn:2019rgj} 
where the one-loop result is written explicitly up to ${\cal O}(\epsilon^4)$.

Let us now discuss how different
quantities scale at subleading level following the general discussion of
Sect.~\ref{sec:leaddel}. The first term of Eq.~(\ref{SUM4}) is the leading term at high energy
discussed in the previous section. The extra $q^2/s$ factor in ${\cal
A}^{(1)}_{SL} $ cancels the Coulomb pole in $A^{(0)}_L$ and, as a result,
we find, after Fourier transforming:
\begin{equation}
  \label{eq:scalingsl}
\left(\frac{i A^{(1)}}{2s}\right)_{SL} \qquad \Rightarrow \qquad G^2 s b^{-2 + 4\epsilon} \sim \left(\frac{R}{b}\right)^{2(1-2 \epsilon)}.
\end{equation}
Note that, in agreement with our general discussion in
Sect.~\ref{sec:leaddel}, we obtain a contribution which, unlike the one of
(\ref{scaling}), does not contain an $\hbar^{-1}$ factor.  For the purpose
of this paper it is enough to carry out the general discussion up to and
including the three-loop order. We have already mentioned the tree and
one-loop order.  In the latter case \eqref{eq:scalingsl} represents the
first term in the expansion of $\Delta$ that appears in \eqref{intro1}.

At two loops, we have the  following hierarchy of contributions 
\begin{eqnarray}
  \label{eq:scaling2l}
\left(\frac{i A^{(2)}}{2s}\right) \qquad\Rightarrow\qquad (\delta_0)^3 &\sim&   \left(\frac{b \sqrt{s}}{\hbar} \left(\frac{R}{b}\right)^{1-2\epsilon}\right)^3~~;~~ 
\nonumber\\
(\delta_0 \Delta_1) &\sim& \delta_2 \sim   \frac{b \sqrt{s}}{\hbar} \left(\frac{R}{b} \right)^{3-6\epsilon}
\end{eqnarray}
and similarly at three loops:
\begin{eqnarray}
  \label{eq:scaling3l}
\left(\frac{i A^{(3)}}{2s}\right) \qquad\Rightarrow\qquad (\delta_0)^4 &\sim&   \left(\frac{b \sqrt{s}}{\hbar} \left(\frac{R}{b}\right)^{1-2\epsilon}\right)^4~~;~~ \nonumber \\
(\delta_0)^2 \Delta_1 &\sim& (\delta_0 \delta_2) \sim   \left(\frac{b \sqrt{s}}{\hbar}\right)^2 \left(\frac{R}{b} \right)^{4(1-2\epsilon)}~~;~~  
\nonumber\\
\Delta_3 &\sim&  \left(\frac{R}{b} \right)^{4(1-2\epsilon)}\; ,
\end{eqnarray}
where $\Delta_3$ is the next term in the expansion of $\Delta$.

Note that at two loops we expect (besides exponentiation of $\delta_0$)
a new classical contribution to the eikonal phase $ \delta_2$,
while at three loops (as it was already the case for one loop) no new
classical contribution is expected. On the other hand, at three loops
the ${\cal{O}}(\hbar^{-4})$ and ${\cal{O}}(\hbar^{-2})$ contributions
should properly reconstruct the relevant terms in \eqref{intro1}.

As already mentioned, for a scattering involving massless particles, the
next-to-leading correction to ${\cal{A}}^{(1)}$ is two powers of centre of
mass energy down with respect to the leading contribution. Thus we do not
have corrections that scale as $(R/b)^{1-4\epsilon} R \sqrt{s}/\hbar$ and would
provide a classical contribution $\delta_1$  entering in the full eikonal
(this is known to be present for the scattering of massive particles,
see e.g. \cite{Luna:2016idw,Bern:2019nnu,KoemansCollado:2019ggb}).
Instead from the subleading part of ${\cal{A}}^{(1)}_{SL}$ we obtain
the first contribution to $\Delta$
\begin{equation}
 2 i \Delta_1 = \int \frac{d^{D-2} q}{(2\pi\hbar)^{D-2}} {\rm e}^{ibq/\hbar}  \frac{i A_L^{(0)}}{2s} \alpha_G {\cal{A}}^{(1)}_{SL}\;.
\label{SUM6}
\end{equation}
By using~\eqref{SUM4} we obtain both the real and the imaginary parts of $\Delta_1$.  
Using the formulas for the Fourier transforms in Appendix \ref{AppB} we get:
\begin{subequations}
  \label{eq:Delta1}
\begin{align}
  Re (2\Delta_1) 
  & = 
\frac{4 G^2 s}{\pi b^2} \left( \pi b^2 \right)^{2\epsilon}
(1+2\epsilon)\Gamma^2 (1-\epsilon) \left[ -\log\left(\frac{s b^2}{4\hbar^2}\right) + H(\epsilon) + \psi(1-2\epsilon) + \psi(\epsilon) \right]\;,
\label{SUM6a} \\
  Im (2\Delta_1)
  & = 
\frac{4 G^2 s}{ b^2} \left( \pi b^2 \right)^{2\epsilon}
 (1+\epsilon)\Gamma^2 (1-\epsilon)\;.
\label{SUM7}
\end{align}
\end{subequations}

Note that, while $Im (2\Delta_1)$ is infrared-finite,  $Re (2\Delta_1)$ is
not since $H(\epsilon) \sim 2 \epsilon^{-1}$.  This may look surprising
at first. In fact, from~\eqref{intro1}, $Re (2\Delta_1)$ appears
to multiply the $S$-matrix by a phase while $Im (2\Delta_1)$  changes
its modulus.  However, if we look at things in terms of the $T$-matrix
($T \equiv -i(S-1) = (A^{(0)} + A^{(1)} + \dots)$), $Im (2\Delta_1)$
comes from (the Fourier transform of) a correction to the phase of
$A^{(0)}$, while $Re (2\Delta_1)$ comes from a negative and infrared
singular correction to its modulus.

More quantitatively, using the small-$\epsilon$ limit of
${\cal{A}}^{(1)}_{SL}$ from~\eqref{SUM4}, the (singular part of the)
one-loop suppression of the  elastic cross section reads:
\begin{equation}
\label{vsuppr}
\sigma_{{\rm el}}^{(1)} \sim \sigma_{{\rm el}}^{(0)} \left(1 + 2 \frac{G q^2}{\pi \epsilon\hbar} ( \log (s/q^2) +1) \right)~ ; ~\sigma_{{\rm el}}^{(0)} \sim |A_0|^2 \, ,
\end{equation}
and is exactly compensated by the cross section for single-soft-graviton
emission. Indeed, the latter is given in terms of  $\sigma_{{\rm
el}}^{(0)}$ by
\begin{equation}
\label{rdivdiff}
 \frac{1}{ \sigma_{{\rm el}}^{(0)}} \frac{d \sigma_{{\rm inelastic}}}{d \omega} \rightarrow  \frac{4G}{\pi \omega\hbar}(s \log s + t \log(-t) + u \log(-u)) \omega^{-2 \epsilon}\, ,
\end{equation} 
which is nothing but the well known (see e.g.~\cite{Addazi:2019mjh}) $D=4$
expression corrected (up to non-singular terms for $\epsilon \rightarrow
0$) in order to account for $D= (4- 2\epsilon)$-dimensional phase space.
Taking the small $q^2/s$ limit of~\eqref{rdivdiff} and integrating it over
$\omega$  leads to the (positive) infrared singular inelastic contribution
\begin{equation}
 \sigma_{{\rm inelastic}} = \int_0 d\omega   \frac{d \sigma}{d \omega}  \sim -\frac {1}{2 \epsilon} \frac{4G q^2}{\pi\hbar} ( \log (s/q^2) +1) \sigma_{{\rm el}}^{(0)}\, ,
  \end{equation}
which exactly cancels the singularity in~\eqref{vsuppr}.

After this digression, we  now use the results~\eqref{SUM6a},\eqref{SUM7}
in~\eqref{intro1}, expand the right-hand side up to order $G^3$, and
compare it with the eikonal expansion of the two-loop amplitude $A^{(2)}$
up to subleading level in the eikonal limit. As we discussed in the
previous section, the highest power of $s$ is entirely reproduced by
the exponentiation of $\delta_0$, so we focus on the next subleading
term, which is of order $G^3 s^2 t$ and yields the  first correction
$\delta_2$ to the leading eikonal $\delta_0$, so the full classical
eikonal $\delta$ is
\begin{equation}
  \label{eq:eikexp}
  \delta = \sum_{n=0}^\infty \delta_{2n}\;,~~\mbox{where}~~~~~ \delta_{2n} \sim \frac{R \sqrt{s}}{\hbar} \left(\frac{R}{b }\right)^{2n(1-2\epsilon)-2 \epsilon} \;.
\end{equation}
By using this in the perturbative expansion of~\eqref{intro1} we can derive $\delta_2$
\begin{align}
  \label{eq:d2deta}
  \frac{A^{(0)}_L}{2s} \alpha_G^2 Re {\cal A}_{SL}^{(2)} & = \int\! { d^{D-2} b}~  {\rm e}^{-ibq/\hbar} \left[- Im(2 \Delta_1) (2 \delta_0) + Re (2 \delta_2) \right]\;,
  \\ \label{eq:d2detb}
  \frac{A^{(0)}_L}{2s} \alpha_G^2  Im {\cal A}_{SL}^{(2)}  & = \int\! { d^{D-2} b} ~{\rm e}^{-ibq/\hbar} \left[Re(2\Delta_1) (2 \delta_0) + Im (2 \delta_2) \right]  \;.
\end{align}
Since we do not have an expression for the two-loop amplitude that is
exact in $\epsilon$ at this order, we are not able to determine the
all-$\epsilon$ form of $\delta_2$.  By using the results expanded around
$\epsilon=0$ of~\cite{BoucherVeronneau:2011qv,Henn:2019rgj} we checked
that~\eqref{eq:d2deta} is consistent with the following expression for
$Re(2 \delta_2)$
\begin{equation}
  Re ( 2 \delta_2 ) = 
\frac{4 G^3 s^2}{\hbar b^2} \left(\pi b^2\right)^{3\epsilon}
\Gamma^3 (1-\epsilon) \left(1+6 \zeta_3 \epsilon^2+\ldots\right)\;.
\label{SUM10}
\end{equation}
In the language of~\eqref{LE3}, this result contains both the contribution
from the exponentiation of IR divergences and that from the remainder
function $F_2$.  The first contribution can be calculated exactly in
$\epsilon$ by using~\eqref{SUM4} and one obtains
\begin{eqnarray}
  \label{EE11v1}
Re (2\delta_2)_{\rm expon} & = &  \frac{4   G^3 s^2 (\pi b^2)^{3\epsilon}
}{\epsilon  \hbar b^2} \left[  B^2(\epsilon) \left( \frac{2(1+\epsilon)
\Gamma (1 - 3 \epsilon)}{\Gamma (1+2 \epsilon)} \right. \right. \\
\nonumber &-& \left. \left.  \frac{(1+2 \epsilon) \Gamma (1-2
\epsilon)}{\Gamma (2+\epsilon)}\frac{\sin \pi \epsilon}{\pi \epsilon}
\left( \frac{s b^2}{4\hbar^2}\right)^{-\epsilon} \right)   - (1+\epsilon)
\Gamma^3 (1- \epsilon) \right] \;.  \end{eqnarray} 
By comparing the
$\epsilon$ expansion of this result with~\eqref{SUM10}, which does
not contain any $\log s$ terms, it is natural to guess that the contribution
of the remainder function should combine with the part proportional to
$B^2(\epsilon)$ in~\eqref{EE11v1}, slightly modifying the normalisation
of the first term and cancelling the contribution of the next term
proportional to $s^{-\epsilon}$. So we can guess a closed  form for the
last two factors\footnote{The discussion in section~\ref{sec:allorders}
uses only on the ${\cal O}(\epsilon)$ part of $\delta_2$ and so it
does not rely on this guess nor on~\eqref{eq:guess2}.} in~\eqref{SUM10}
\begin{eqnarray}
 \label{eq:guess1}
\Gamma^3 (1-\epsilon) \left(1+6 \zeta_3 \epsilon^2+\ldots\right)& =&
\frac{1 }{\epsilon}\left( B^2(\epsilon) \frac{ (1+2 \epsilon) \Gamma(1-3
\epsilon)}{\Gamma (1+ 2 \epsilon) }- (1+\epsilon) \Gamma^3 (1-\epsilon)
\right)  \nonumber \\ &=&\frac{\Gamma^3 (1-\epsilon)}{\epsilon} \left(
\frac{1+2\epsilon}{G^{(2)} (\epsilon)} - (1+\epsilon) \right)  \;,
\end{eqnarray} 
where $G^{(2)} (\epsilon)$ is defined in \eqref{eq:Gedef}
and the $(1+\epsilon)$ term in the second line comes from subtracting the
$-2\delta_0 2Im \Delta_1$ piece in~\eqref{eq:d2deta}, which again can
be derived exactly in $\epsilon$. The contribution from the remainder
function is then qualitatively similar, but quantitatively different
from the one coming from exponentiation of the one-loop result and can be
derived by comparing~\eqref{EE11v1} and~\eqref{SUM10} after including the
guess~\eqref{eq:guess1} 
\begin{eqnarray} Re (2\delta_2)_{\rm remainder} 
& = &  \frac{4   G^3 s^2 (\pi b^2)^{3\epsilon} }{\epsilon \hbar b^2} \left[  B^2(\epsilon)
\left(- \frac{ \Gamma (1 - 3 \epsilon)}{\Gamma (1+2 \epsilon)}
\right. \right. \nonumber \\ &+ & \left. \left.  \frac{(1+2 \epsilon)
\Gamma (1-2 \epsilon)}{\Gamma (2+\epsilon)}\frac{\sin \pi \epsilon}{\pi
\epsilon} \left( \frac{s b^2}{4\hbar^2}\right)^{-\epsilon} \right)
\right] \, .  \label{EE11} \end{eqnarray}
The need for such a
complicated remainder can be understood to follow from a very physical
requirement. Since the derivative of $Re\,\delta_2$ w.r.t. $b$ gives a
correction to the physical deflection angle, we can reasonably require
that it should have a finite classical limit. However, for dimensional
reasons, any dependence on $ b^2 s \sim J^2$ needs to be interpreted as
a dependence from $\frac{J^2}{\hbar^2}$ which would lead to a  divergent
deflection angle  in the classical ($\hbar \rightarrow 0$) limit for
generic values of $D$. As a consequence, the remainder's contribution must
have the correct $b^2s$-dependent piece as given in~\eqref{EE11}. This,
however, is not enough since that piece  is infrared divergent while the
remainder, by its definition, is not. The additional term $- \Gamma (1 -
3 \epsilon)/\Gamma (1+2 \epsilon)$  fixes (although not in a unique way)
this last problem. 

We thus learn  that the separation of the full amplitude into
an exponential piece and a remainder is hiding a simple physical
property. The remainder has to be a complicated function of $b^2 s$
so that the full amplitude does not depend on it! Or, turning things
around, we can say that a simple physical requirement determines a very
non trivial structure for the remainder (in analogy with the consequences
of exponentiation discussed in \cite{DiVecchia:2019myk}).

Turning now to  $Im (\delta_2)$ we find, using~\eqref{eq:d2detb} and
again the results of \cite{BoucherVeronneau:2011qv,Henn:2019rgj}:
\begin{align}
  \nonumber Im( 2 \delta_2 ) = &
\,-\, \frac{ 4 G^3 s^2}{\pi \hbar b^2} \left( \pi b^2 \right)^{3\epsilon} \frac{(1-2\epsilon) \Gamma^3
(1-\epsilon)}{\epsilon} \Bigg[\Big(1-12 \epsilon^3 \zeta_3+\ldots \Big)
  \log\left(e^{2\gamma_E}\frac{s\, b^2}{4\hbar^2}\right) \\ + & \left(1
  - 3 \zeta_2 \epsilon + (-23 \zeta_3 - 32 \zeta_2) \epsilon^2 + (-167
  \zeta_4 - 160 \zeta_3 - 64 \zeta_2) \epsilon^3 +\ldots\right)\Bigg]\;.
  \label{SUM11m}
\end{align} 
A possible guess for the factor in the first line of~\eqref{SUM11m}
that uses the same function $G^{(2)}$ encountered before is as follows
\begin{equation}
  \label{eq:guess2}
   \Big(1-12 \epsilon^3 \zeta_3+\ldots \Big) = \left(3-\frac{2 \Gamma (1-3 \epsilon) \Gamma  (1-\epsilon) \Gamma^2 (\epsilon+1)}{\Gamma^2 (1-2 \epsilon) \Gamma (2 \epsilon+1)}\right)= 3 - \frac{2}{G^{(2)} (\epsilon)}\;,
 \end{equation}
while we do not currently have a guess for the factor in the second line of~\eqref{SUM11m}.
 
Again this should match the result from the exponentiation and remainder
contributions.  A long but straightforward calculation gives
\begin{eqnarray} \label{EE42}
Im (2 \delta_2)_{\rm expon} 
& = & \frac{8G^3 s^2}{\pi  \hbar b^2 } 
(\pi b^2)^{3\epsilon} 
\frac{ (1+2\epsilon)  \Gamma^3 (1-\epsilon)}{\epsilon^2}
\Bigg\{ \frac{Y}{2} 
+  \frac{\Gamma (1-\epsilon) \Gamma^2 (1+\epsilon)}{\Gamma (1-2\epsilon)} 
\\ \nonumber
&\times&
\left[ \left( \frac{s b^2}{4\hbar^2} \right)^{-\epsilon} \frac{\cos^2 \frac{\pi \epsilon}{2}}{\Gamma (\epsilon+2)} 
- \frac{\Gamma (1-3\epsilon)}{\Gamma (1-2\epsilon) \Gamma (1+2 \epsilon)} X  \right] \Bigg\} \, ,
\end{eqnarray}
where we have defined:
\begin{eqnarray}
X &=& \epsilon \left( \pi \cot \pi \epsilon - 
\log \left(\frac{s b^2}{4\hbar^2}\right) 
-1 +\gamma_E + \psi(-\epsilon) + \psi (2\epsilon) + \psi (1-3\epsilon)\right) \label{EE41a} \\
Y &=& \epsilon \left( \pi \cot \pi \epsilon - 
\log \left(\frac{s b^2}{4\hbar^2} \right)
-1 +\gamma_E + \psi(-\epsilon) + \psi (\epsilon) + \psi (1-2\epsilon)\right) \;.
\label{EE41b}
\end{eqnarray}
On the other hand the leading term in the $\epsilon$-expansion of the
remainder can be extracted from the imaginary part of~\eqref{eq:F2e0}
and gives:
\begin{equation}
  Im (2 \delta_2)_{\rm remainder} 
     = -\frac{4 s^2 G^3 }{\pi \hbar b^2}
\left[  \log^2 
\left( \frac{s b^2}{4\hbar^2} e^{2\gamma_E} \right) 
- 2 \log 
\left( \frac{s b^2}{4\hbar^2} e^{2\gamma_E} \right) 
     -2\left(1+  \frac{\pi^2}{3}\right) \right] + {\cal O}(\epsilon)\,.
\label{EE53}
\end{equation}
One can check that such a remainder gives agreement
with~\eqref{SUM11m} at $\epsilon =0$.  On the other hand, also this
time the remainder's contribution has to be highly non trivial in
order to reconcile~\eqref{EE42} with~\eqref{SUM11m} at finite
$\epsilon$. In particular, the power of the $b^2 s$ dependence
of~\eqref{EE42} has to be cancelled by the remainder leaving just a
single (and singular\footnote{ Note that, for $Im (\delta_2)$, there
  is nothing wrong with IR divergences since they are related to
  bremsstrahlung processes.}) $ \log \left( sb^2/4 \hbar^2 \right) $
like those appearing in $X$ and $Y$ of~\eqref{EE41a}
and~\eqref{EE41b}.  Once more this shows that the separation of the
full amplitude into an exponential of the one-loop result and a
remainder hides some simple feature of the impact-parameter result.

Let us now discuss some physical consequences of the above results.
Notice first that the term of order $\epsilon^0$  in~\eqref{SUM10}
\begin{eqnarray}
\lim_{\epsilon \rightarrow 0} Re (2 \delta_2) = \frac{4 G^3 s^2}{\hbar b^2}
\label{SUM11}
\end{eqnarray}
is identical to Eq. (5.26) of~\cite{Amati:1990xe} where this quantity
has been computed  for pure gravity.\footnote{Before comparing this
result to those obtained by other methods, one should be careful about
the relation between $b$ and the actual total angular momentum $J$ in
the process. The calculation of the deflection angle to this 
subleading level is sensitive to this precise relation.} Since we have obtained it
for ${\cal{N}}=8$ supergravity, this appears to indicate that classical
quantities, such as $Re \delta_2$, are related only to large-distance
physics and are therefore independent of the ultraviolet behavior of the
microscopic theory and thus universal.\footnote{This universality has
been known for sometime \cite{BoucherVeronneau:2011qv} for the leading
eikonal. We have been informed by Parra-Martinez that it has also
been checked at this subleading level for $4 \le {\cal{N}}  \le 8$. A
first hint for such universality goes back to \cite{Bellini:1992eb}.}
We checked~\eqref{SUM10} up to order $\epsilon^2$ by verifying
that~\eqref{eq:d2deta} reproduces the results of the two-loop amplitude
in dimensional regularisation~\cite{BoucherVeronneau:2011qv,Henn:2019rgj}.

Turning to $Im \delta_2$,  a few interesting properties of
Eqs.~\eqref{SUM10} and~\eqref{SUM11m} should be stressed: 
\begin{itemize}
 \item $Im(\delta_2)$ and $Re(\delta_2)$ both scale like
$G^3 s^2 \hbar^{-1} (b^2)^{-1+3 \epsilon}$.
 \item Unlike $Re(\delta_2)$, which is regular for $\epsilon \rightarrow
 0$, $Im(\delta_2)$ is singular.  
\item Nonetheless, $Im(\delta_2)$
 does not have ${\cal O}(\epsilon^{-2})$ singularities and its ${\cal
 O}(\epsilon^{-1})$ term multiplies just the
combination $ \log\left(\frac{s\, b^2 }{4 \hbar^2} {\rm e}^{2
\gamma_E}\right) +1$.
 \item At ${\cal O}(\epsilon^{0})$,  $Im(\delta_2)$ develops a term
 proportional to $ \log\left(\frac{s\, b^2}{4\hbar^2}\right) \log(
 \pi b^2)$.  \end{itemize} 
This is also in line with the findings  of
Ref.~\cite{Amati:1990xe} (see Eq. (5.26) there) for the pure gravity
case. While $Re(\delta_2)$ is directly related to a physical observable,
the deflection angle, $Im(\delta_2)$ is related to gravitational
bremsstrahlung with its well-known infrared divergences.

It is also amusing to compare $Im(\delta_2)$ with the Fourier transform
of the imaginary part of the  full two-loop amplitude. The latter can
be found either by adding to~\eqref{SUM11m} the known contribution
of $2\delta_0  Re(2\Delta_1)$ according to Eq.~\eqref{eq:d2detb},
or by simply starting from the expression given in Appendix~\ref{AppC}
(Eq.~\eqref{eq:a2L}).  The result, up to terms that vanish for $\epsilon
\rightarrow 0$, can be expressed in a particularly simple form:
\begin{equation}
  \label{FTImA2}
\int \frac{d^{D-2} q}{(2\pi\hbar)^{D-2}}
{\rm e}^{ibq/\hbar}
  \frac{Im A^{(2)}_{SL} }{2s}
=  -  \frac{4 G^3 s^2}{\pi\hbar b^2}\left[ \frac{1}{\epsilon^2} \left(
\pi b^2 \exp(\gamma_E + 2/3) \right)^{3\epsilon} - 4 \log\left(\frac{s
b^2 }{4\hbar^2}\right)+ C\right]\,,
   \end{equation}
where $C =-8(\gamma_E + 3/4) -\frac{11}{12}\pi^2 = -19.6649 \dots$
Note that, unlike $Im(\delta_2)$, this quantity does have an ${\cal
  O}(\epsilon^{-2})$ singularity. However this, as well as an ${\cal
  O}(\epsilon^{-1})$ singularity, only concerns terms
involving $\log (b^2)$ and $\log^2 (b^2)$ and {\it
not} $\log(s b^2)$.  The latter only occurs at ${\cal
O}(\epsilon^{0})$. The presence of a double pole in the amplitude
itself arises from the known exponentiation of IR singularities in
gravity~\cite{Weinberg:1965nx,Naculich:2011ry,White:2011yy,Akhoury:2011kq}.
Denoting the ${\cal O}(\epsilon^{m})$ part of the $\ell$-loop amplitude by
${\cal A}^{(\ell,m)}$, one has
\begin{equation}
{\cal A}^{(2,-2)}=\frac{1}{2}\left[{\cal A}^{(1,-1)}\right]^2.
\end{equation}
From Eq.~\eqref{SUM4}, one finds
\begin{equation}
{\cal A}^{(1,-1)}=-i\pi s+q^2\left[\log\left(\frac{s}{q^2}\right)+1\right],
\end{equation}
and thus 
\begin{equation}
Im{\cal A}^{(2,-2)}=\pi q^2 s\log(q^2)+\ldots,
\end{equation}
in agreement with Eq.~\eqref{eq:a2L}, where the ellipsis denotes terms
analytic in $q^2$. Note that this is not inconsistent with the lack of
a double $\epsilon$ pole in $Im(\delta_2)$: the latter is in the
logarithm of the amplitude, and thus does not contain that part of the
two-loop amplitude which results from the exponentiation of
lower-order results. 

\section{Comparing the two exponentiations}
\label{sec:allorders} 

In ref.~\cite{DiVecchia:2019myk} (and reviewed in sec.~\ \ref{sec:leadexp})
we told the tale of how the exponentiations 
in impact-parameter space and in momentum space are related
for the leading high-energy terms of the amplitude.
These exponentiations differ in significant respects:
in impact parameter space, the exponentiation 
starts at tree level with the eikonal phase, 
and the eikonal phase is IR-divergent. 
In momentum space, the tree-level amplitude is IR-finite,
and the exponentiation starts with the IR-divergent
one-loop amplitude. 
Nevertheless, the first type of exponentiation implies the second, 
up to an IR-finite
correction factor (given by the expression 
$G^{(\ell)}(\epsilon)$ in Eq.~\eqref{LE14}) 
which determines the leading-order contribution 
to the remainder function.

In this section, we relate a similar connection between
impact-parameter space and momentum space amplitudes 
at the first subleading level. 
That is, we show that the proposed extension (\ref{intro1})
of the eikonal amplitude 
\begin{equation}
\frac{i A (k_i,\ldots)}{2s} \simeq \hat{A}^{(0)} (k_i,\ldots) \int\! { d^{D-2} b} ~{\rm e}^{-ibq/\hbar} \, \left[\Big(1+2i\Delta(s,b) \Big)\, {\rm e}^{2i\delta (s, b)} -1\right] 
\label{eq:proposed}
\end{equation}
agrees with the expected exponentiation in momentum space at first
subleading level in $q^2/s$, to at least the first two orders in the
Laurent expansion in $\epsilon$.

The leading and first subleading contributions are given by
\begin{align} 
\frac{i A_L}{2s} 
&= \hat{A}^{(0)} (k_i,\ldots) 
\int\! { d^{D-2} b} ~{\rm e}^{-ibq/\hbar} \, 
\left( {\rm e}^{2i\delta_0}  - 1 \right) \;, 
\label{eq:L}
\\
\frac{i A_{SL} }{2s} 
&\simeq \hat{A}^{(0)} (k_i,\ldots) 
\int\! { d^{D-2} b} ~{\rm e}^{-ibq/\hbar} \, 
\left( 2i \Delta_1 \sum_{\ell=1}^\infty \frac{(2i \delta_0)^{\ell-1}}{(\ell-1)!} + 2i \delta_2 \sum_{\ell=2}^\infty \frac{(2i \delta_0)^{\ell-2}}{(\ell-2)!}
\right)\;.
\label{eq:SL}
\end{align} 
We have already considered the leading contribution \eqref{eq:L} 
in sec.~\ref{sec:leadexp}.
To compute the subleading contribution \eqref{eq:SL},  
we use
\begin{equation}
2 i \delta_0 
= \, -\, { i G s   \over \epsilon \hbar} 
\Gamma(1-\epsilon)\left(  \pi  b^2 \right)^{\epsilon }
\end{equation}
together with the expressions for 
$\Delta_1$ and $\delta_2$ obtained in sec.~\ref{sec:subleaddel} 
\begin{align}
 2 i \Delta_1
&= 
{4 i G^2 s \Gamma^2(1-\epsilon)  \over  (\pi b^2)^{1-2\epsilon} }
\left( (1 + 2 \epsilon) \left[ -\log\left(s b^2 \over 4\hbar^2\right) 
+ H(\epsilon) + \psi(1-2\epsilon) + \psi(\epsilon)\right]  + i \pi (1+\epsilon) \right) \;,
\nonumber \\ \label{eq:dDsum}
2 i \delta_2
&=  {4G^3 s^2 \Gamma^3(1-\epsilon)
\over  \epsilon \hbar( \pi b^2)^{1-3\epsilon}} 
\left(  D_1(\epsilon) \log\left(e^{2\gamma_E} {s b^2 \over 4\hbar^2}\right)  + D_2(\epsilon) \right)\;,
\end{align}
where 
\begin{align}
D_1(\epsilon) & = (1-2\epsilon) \left( 3 - \frac{2}{G^{(2)}(\epsilon)} \right)  
\nonumber \\
 &
= 1 - 2 \epsilon - 12 \zeta_3 \epsilon^3 + {\cal O} (\epsilon^4)  \,,
\nonumber\\
D_2(\epsilon) &= 
(1-2 \epsilon) L(\epsilon)  +i\pi \left( \frac{1+2\epsilon}{G^{(2)} (\epsilon) } -1-\epsilon \right)
\nonumber \\
 &
= (1-2\epsilon) L(\epsilon) + i \pi \epsilon \left( 1 + 6 \zeta_3 \epsilon^2 \right)  
+ {\cal O} (\epsilon^4) 
\label{eq:Ddef}
\end{align}
and
\begin{equation}
L(\epsilon) = 1 - 3 \zeta_2 \epsilon + (-23 \zeta_3 - 32 \zeta_2) \epsilon^2 
+ (-167 \zeta_4 - 160 \zeta_3 - 64 \zeta_2) \epsilon^3  + {\cal{O}}(\epsilon^4)\, ,
\label{Lepsilon}
\end{equation}
where the terms with $G^{(2)} (\epsilon)$ are possible guesses to
any order in $\epsilon$ of quantities that are known only up to order
$\epsilon^3$.  Using
\begin{equation}
\frac{i A^{(0)}}{2s} 
= \frac{i A_L^{(0)}}{2s} \hat{A}^{(0)} (k_i,\ldots) 
= \frac{4 \pi i G \hbar s}{q^2 } \hat{A}^{(0)} (k_i,\ldots) 
\end{equation}
together with eqs.~\eqref{B1bis} and \eqref{logb},
the computation of \eqref{eq:SL} is straightforward.
The $\ell$-loop subleading contribution is  
\begin{align}
\frac{i A^{(\ell)}_{SL} }{2s} 
&\simeq  
\frac{i A^{(0)} }{2s} 
 {\alpha_G^\ell  \over \ell!} 
 \left[  {- i \pi s \over \epsilon (q^2)^\epsilon} 
\right]^\ell 
\frac{iq^2}{\pi s}
G^{(\ell)} (\epsilon) 
\label{eq:subleading}
\\
&\times
\Bigg\{ (1 + 2 \epsilon) \left[ -\log
\left(s \over q^2\right) 
+ H(\epsilon) 
+ \psi(1-2\epsilon) + \psi(\epsilon)
- \psi(1 - (\ell+1)\epsilon) - \psi (\ell \epsilon)
\right]  
\nonumber\\
& + i \pi (1+\epsilon) 
+ (\ell-1) D_1(\epsilon) 
\left[ \log
\left(  e^{2 \gamma_E} {s \over q^2}\right) 
+ \psi(1 - (\ell+1)\epsilon) + \psi (\ell \epsilon) \right] 
+ ( \ell-1  ) D_2(\epsilon) 
\Bigg\}\;.
\nonumber
\end{align}
The divergent terms in this expression should match those arising from
the IR exponentiation in~\eqref{LE3}. We start by considering the first
two terms in the $\epsilon$ expansion where one can neglect the remainder
functions appearing in~\eqref{LE3}. Then in a separate subsection we
consider the third and the fourth terms in the $\epsilon$ expansion:
the third order depends on the finite part of $F^{(2)}$, while the
fourth one receives contributions also from the ${\cal O}(\epsilon)$
term in $F^{(2)}$ and the finite part of $F^{(3)}$.

\subsection{The first two leading orders in $\epsilon$ at $\ell$-loop order}
\label{sec:elm1} 

As mentioned previously, 
the eikonal expression \eqref{eq:proposed} is only meant 
to capture the 
non-analytic contributions to the 
momentum space amplitude as $q^2 \to 0$.  
Additional polynomial terms in $q^2$ will Fourier transform 
to give $\delta^{(d-2)} (b)$ function terms (or derivatives thereof)
in impact parameter space.
To identify all non-analytic terms  in \eqref{eq:subleading},  
we must expand
$(q^2)^{-\ell  \epsilon} = \exp[ - \ell \epsilon \log(q^2) ]$ 
in $\epsilon$.
In addition we use 
$G^{(\ell)}(\epsilon) = 1 + {\cal O} (\epsilon^3) $
and Laurent expand the functions
\begin{align}
 H(\epsilon) 
+ \psi(1-2\epsilon) + \psi(\epsilon)
- \psi(1 - (\ell+1)\epsilon) - \psi (\ell \epsilon)
&=
\left(\ell+1\over \ell\right) {1 \over \epsilon} - 1 + {\cal O}(\epsilon) \;,
\nonumber\\
 \psi(1 - (\ell+1)\epsilon) + \psi (\ell \epsilon) 
&=
 \, -\, {1 \over \ell \epsilon} - 2 \gamma_E + {\cal O}(\epsilon)  \;.
\end{align}
Dropping all the terms in \eqref{eq:subleading}
that have no $\log^n (q^2)$-dependence, we obtain
\begin{align}
\frac{i A^{(\ell)}_{SL} }{2s} 
&\simeq  
\frac{i A^{(0)} }{2s} 
 {\alpha_G^\ell  \over \ell!} 
 \left(  {- i \pi s \over \epsilon}  \right)^\ell
\frac{iq^2}{\pi s}
\bigg\{
-\ell \log \left(q^2 \right) 
\nonumber\\
&
+ \epsilon \bigg[  
{\textstyle{1 \over 2}}
    {\ell(\ell-1)} \big[ D_1(0)+1\big]  \log^2  \left(q^2  \right) 
+ \ell \big[   (1-\ell) D_1(0) + 1 \big] 
\log \left(s \right) \log \left(q^2  \right)
\nonumber\\
&
+ \ell  \big[ 
 ( 1-\ell)  D_2(0) -1 \big] 
  \log \left(q^2  \right)
-i \pi \ell \log \left(q^2  \right) 
\bigg]
+ {\cal O}(\epsilon^2)
\bigg\} \,.
\label{eq:qdep}
\end{align}
By noting that \eqref{eq:Ddef} implies 
$D_1(0)  = D_2(0) = 1$ 
(where the imaginary part of $D_2(\epsilon)$ 
only begins at ${\cal O}(\epsilon)$),
we obtain all the nonanalytic subleading terms
through ${\cal O} (1/\epsilon^{\ell-1})$:
\begin{align}
\frac{i A^{(\ell)}_{SL} }{2s} 
&\simeq  
\frac{i A^{(0)} }{2s} 
 {\alpha_G^\ell  \over \ell!} 
 \left(  {- i \pi s \over \epsilon}  \right)^\ell
\frac{iq^2}{\pi s}
\bigg\{
-\ell \log \left(q^2 \right) 
\nonumber\\
&
\qquad + \epsilon \bigg[  
 {\ell(\ell-1)}   \log^2  \left(q^2  \right) 
-  \ell (\ell-2) 
\log \left(s \right) \log \left(q^2  \right)
\nonumber\\
&
\qquad -  \ell^2 \log \left(q^2  \right)
-i \pi \ell \log \left(q^2  \right) 
\bigg]
\bigg\}
+ {\cal O} ( 1 / \epsilon^{\ell-2} ) \;.
\label{eq:subeik}
\end{align}
Now let us check this against the expected exponentiation
in momentum space
\begin{equation}
{iA \over 2s}  = {i A^{(0)}  \over 2s}
\exp \Big( \alpha_G {\cal{A}}^{(1)} \Big) 
\exp \left( \sum_{\ell=2}^{\infty} \alpha_G^{\ell} F^{(\ell)}\right)\;,
\end{equation}
where 
\begin{equation}
{\cal{A}}^{(1)} 
= 
{1 \over \epsilon (q^2)^\epsilon}
\left[ - i \pi s + q^2  
\left(
\log\left( s \over q^2 \right)  
+1\right) 
\right]
 + {q^2 \over (q^2)^\epsilon}
\left[ - 
\log^2 \left( s \over q^2 \right)  
+ i \pi 
\log \left( s \over q^2 \right)  
\right]
 + {\cal O} (\epsilon)\;.
\label{eq:oneloop}
\end{equation}
Since the remainder function $F^{(\ell)}$  is IR-finite and only begins at two-loop order,
the first two terms in the Laurent expansion of the $\ell$-loop amplitude
are completely dictated  by the one-loop amplitude
\begin{equation}
{iA^{(\ell)} \over 2s}   = {i A^{(0)}  \over 2s}
{\alpha_G^\ell \over \ell!}  \left( {\cal{A}}^{(1)} \right)^\ell
+ {\cal O} ( 1 /  \epsilon^{\ell-2} ) \,.
\label{eq:ellloop}
\end{equation}
Substituting Eq.~\eqref{eq:oneloop} into \eqref{eq:ellloop},
we obtain for the leading level $\ell$-loop amplitude
\begin{equation}
{iA_L^{(\ell)} \over 2s}   = {i A^{(0)}  \over 2s}
{\alpha_G^\ell \over \ell!} 
\left(  {- i \pi s  \over \epsilon (q^2)^\epsilon}  \right)^\ell 
+ {\cal O}(1/\epsilon^{\ell-2})
\end{equation}
agreeing with the leading level eikonal expression 
\eqref{LE14} to this order in $\epsilon$.
For the subleading level $\ell$-loop amplitude, we get
\begin{align}
{iA_{SL}^{(\ell)} \over 2s}   &= {i A^{(0)}  \over 2s}
{\alpha_G^\ell \over \ell!} 
\left(  {- i \pi s  \over \epsilon (q^2)^\epsilon}  \right)^\ell 
{iq^2 \ell \over s \pi}
\Bigg\{
\log\left( s \over q^2 \right)  
+1
\nonumber\\
& 
 \qquad\qquad\qquad 
+ \epsilon \left[ - 
\log^2 \left( s \over q^2 \right)  
+ i \pi 
\log \left( s \over q^2 \right)  
\right]
\Bigg\}
+ {\cal O}(1/\epsilon^{\ell-2})
\\
&=
 {i A^{(0)}  \over 2s}
{\alpha_G^\ell \over \ell!} 
\left(  {- i \pi s  \over \epsilon}  \right)^\ell 
{iq^2  \over s\pi}
\Bigg\{
  \ell\log\left(s\right) 
+ \ell 
- \ell\log\left(q^2\right) 
\nonumber\\
& 
 \qquad\qquad\qquad 
+ \epsilon \bigg[
- \ell\log^2 \left(s\right) 
+ i \pi \ell\log \left(s\right) 
+\ell(\ell-1) \log^2 \left(q^2\right)
\nonumber\\
&
 \qquad\qquad\qquad 
- \ell(\ell-2) \log \left(s\right) \log \left(q^2\right) 
- \ell^2 \log \left(q^2  \right) 
- i \pi \ell\log \left(q^2\right)
\bigg]
\Bigg\}
+ {\cal O}\left(\frac{1}{\epsilon^{\ell-2}}\right)\,.
\nonumber
\end{align}
Comparing the $\log^n (q^2)$-dependent terms of this expression with \eqref{eq:subeik}
we find perfect agreement.

\subsection{The first four leading orders in 
$\epsilon$ at $\ell$-loop order}
\label{sec:elm3} 

So far we exploited only the knowledge of the one-loop amplitude
in evaluating~\eqref{LE3}, but thanks to the explicit results
of~\cite{Henn:2019rgj} we can extend the comparison between the two
exponentiations at the subleading level in the eikonal limit to the
first four terms in the $\epsilon$ expansion.

Let us start by analysing in some detail the three-loop case. The
leading term of $A^{(3)}/(2s)$ scales as $s^4$ and, as discussed in
section~\ref{sec:leadexp}, it is entirely reproduced by the
exponentiation of $\delta_0$. The subleading contribution
$A_{SL}^{(3)}/(2s)$ scales, after Fourier transform to impact parameter space,
as $ (Gs/\hbar)^2 (R/b)^2 \log^{n-1}(b^2)$ and, as discussed before, we focus on the
long-range contributions, i.e. the terms with $n\geq 1$. From the
scaling above it is clear that such terms grows too quickly with the
energy (and is too singular in the classical limit) to be absorbed in
a contribution $\delta_3$ to the total eikonal or in a contribution
$\Delta_3$ to the prefactor $\Delta$. Thus they must be reproduced by
the leading and the subleading eikonal data, as dictated
by~\eqref{eq:SL}.  Then, by separating the real and the imaginary
parts, we have\footnote{We can write the three-loop consistency
  condition dictated by assuming the eikonal
  exponentiation~\eqref{intro1} in the momentum space language as
  follows:
\begin{equation}
\label{eq:CCcalA}
  \frac{A^{(0)}_L}{2s} \alpha_G^3 {\cal A}^{(3)} = \frac{1}{2} \int\! { d^{D-2} b} {\rm e}^{-ibq/\hbar} \Big[(2 \delta_0)^2 (2 \Delta_1) \Big] + i \int \frac{d^{D-2}k}{(2\pi\hbar)^{D-2}} \frac{(4\pi G\hbar  s)^2}{(q-k)^2 k^2} \alpha_G^2  {\cal A}^{(2)}(s,k^2)\;.
\end{equation}}
\begin{equation}
\frac{A^{(0)}_L}{2s}  \alpha_G^3  Re{\cal A}_{SL}^{(3)}  = \int\! { d^{D-2} b}\, {\rm e}^{-ibq/\hbar}\left[- \frac{1}{2} (2\delta_0)^2 Re (2 \Delta_1)  - (2 \delta_0) Im (2\delta_2) \right]\;
\label{SUM16bis}
\end{equation}
and similarly for the imaginary part 
\begin{equation}
\frac{A^{(0)}_L}{2s} \alpha_G^3 Im{\cal A}_{SL}^{(3)} = \int\! { d^{D-2} b}\, {\rm e}^{-ibq/\hbar}\left[ -\frac{1}{2} (2\delta_0)^2 Im(2 \Delta_1) + (2 \delta_0)  Re(2 \delta_2) \right]\;.
  \label{SUM16}
\end{equation}

The left-hand side of these equations can be extracted from the
full three-loop ${\cal N}=8$ 4-point amplitude recently derived
in~\cite{Henn:2019rgj}.  The relevant terms in the Regge regime up
to the first subleading level in the Regge limit are summarised in
Appendix~\ref{AppC}.  The right-hand side is obtained by using \eqref{LE6}
for $\delta_0$,~\eqref{eq:Delta1} for $\Delta_1$,~\eqref{SUM10}
for $Re(\delta_2)$, and~\eqref{SUM11m} for $Im(\delta_2)$.  The
relation~\eqref{SUM16} is easier to check since $Re(\delta_2)$ is simpler
than $Im(\delta_2)$.  The left-hand side is given by the five imaginary
terms of the subleading ({\rm i.e.} proportional to $s^2$) contribution
in~\eqref{eq:a3L}. We checked that the eikonal exponentiation on the
right-hand side of~\eqref{SUM16} reproduces exactly these terms.

We performed a similar check for~\eqref{SUM16bis}. Now the left-hand side
involves eighteen terms which are the real contributions to the $s^2$ part
of~\eqref{eq:a3L}. The structure of the answer is more complicated and
includes contributions enhanced by a factor of $\log(s)$. By comparing
this result with the prediction on the right-hand side coming from
the eikonal exponentiation we find agreement for all terms but one.
In particular all divergent terms as $\epsilon \to 0$ and all terms
proportional to $\log^n(q^2)$ with $n\geq 2$ match. However by going
all the way down to the lowest order contribution (i.e of ${\cal O}(G^4
s^3/b^2)$ with no $\log s$ enhancement) we find a mismatch, which,
in momentum space, reads:
\begin{equation}
  \label{eq:mismatch}
({\rm lhs}-{\rm rhs})_{Eq.~\eqref{SUM16bis}} 
= \frac{16}{3} {G^4 s^3\over \hbar^2}  \left(3\zeta_3-\pi^2  \right)\log(q^2) \;.
\end{equation}
From~\eqref{eq:mismatch} we see that the mismatch is sensitive to the two-loop contribution proportional to $\epsilon \log q^2$ and to the three-loop contribution proportional to $\log q^2$. 
Suppose one were to modify these terms in the amplitude
\begin{equation}
  \label{eq:A3mod}
  \tilde{\cal A}^{(2)} = {\cal A}^{(2)} +  i \pi \epsilon c_2  s q^2 \log q^2 +\ldots\;, \quad   \qquad
  \tilde{\cal A}^{(3)} = {\cal A}^{(3)} + \pi^2 c_3 s^2 q^2 \log q^2 +\ldots\;,
\end{equation}
where the ${\cal A}$'s on the right-hand side are 
those given in \eqref{eq:a2L} and \eqref{eq:a3L} 
and the dots stand for further analytic contributions or higher order terms in $\epsilon$. 
This would change the remainder functions 
from the ones given in \eqref{eq:F2e0} and \eqref{eq:F3e0} to 
\begin{equation}
  \label{eq:F3mod}
\tilde{F}^{(2)} =  F^{(2)} + ~  i \pi \epsilon c_2  s q^2 \log q^2\;, \quad \qquad  
\tilde{F}^{(3)} =  F^{(3)} +   \pi^2 (c_3-c_2) s^2 q^2 \log q^2\;,
\\ 
 \end{equation}
and the eikonal in \eqref{eq:dDsum} to 
\begin{equation}
  \label{eq:d2mod}
   \tilde{\delta}_2 = \delta_2 - {4 i G^3 s^2 \epsilon\, \Gamma^3(1-\epsilon)
\over \hbar  (1-2 \epsilon) ( \pi b^2)^{1-3\epsilon}} + {\cal O}(\epsilon^2)\;.
\end{equation}
The tilde'd quantities
now satisfy the consistency check~\eqref{SUM16bis},
provided that the parameters appearing in~\eqref{eq:A3mod}
satisfy the constraint
\begin{equation}
  \label{eq:c3c2}
  c_3 = c_2 - \frac{4}{3} \left(3 \zeta_3 - \pi^2\right)\;.
\end{equation}
This modification, however, turns out to be insufficient to cure a mismatch at higher-loop order, as we shall now argue.

We can follow the logic of~\eqref{sec:elm1} and use the first four terms
in the $\epsilon$-expansion of the $\ell$-loop result for~\eqref{eq:SL}
as a check of remainder functions proposed in~\eqref{eq:F3mod}.
The $\ell$-loop eikonal prediction \eqref{eq:SL} for the subleading
amplitude still does not agree with the (IR-divergent) prediction of the
momentum-space exponentiation \eqref{LE3}, even when using the modified
remainder functions \eqref{eq:F3mod}.  Furthermore, this mismatch is
independent of the choice for the residual parameter $c_2$, which is thus
unfixed by these checks.  The mismatch first appears at order $1/\epsilon
^{\ell-3}$ (for $\ell>3$), and has the following pattern:
\begin{equation}
  \label{eq:mismatch2}
   ({\rm lhs}-{\rm rhs})_{Eq.~\eqref{eq:SL}} \sim \frac{i \pi s q^2 \log q^2}{\epsilon^{\ell-3} (\ell-4)!} \;,
\end{equation}
where the proportionality constant is independent of $\ell$ 
and all the quantities are calculated using \eqref{eq:A3mod}--\eqref{eq:c3c2}.
Amazingly, 
the mismatch \eqref{eq:mismatch2} could be avoided for all $\ell$
by the following further redefinition of the three-loop remainder function
\begin{equation}
  \label{eq:F3mod2}
  \hat{F}^{(3)} = \tilde{F}^{(3)} +2  \pi^2 s^2 q^2 \, \frac{\zeta_3}{\epsilon}\;.
\end{equation}
Such a redefinition,
however, is not allowed if all infrared divergences are captured by the exponentiation of the one-loop result 
as assumed in~\eqref{LE3}.

It is difficult to assess the meaning of the few mismatches we
  found when weighed against the large number of successful
checks. One possibility is that factorization can slightly break in
the non-conservative contributions to the amplitude since, by
themselves, they do not carry a physical meaning. If so, one should
check whether some inconsistency is still present after computing a
more physical quantity such as an infrared-finite inclusive cross
section. Another, perhaps more interesting possibility, is that the
two results have different regimes of validity depending on whether
the IR cutoff is the lowest energy scale in the problem or not. We
will add further comments on this point in the final section.


\section{The $D=4$ eikonal using a momentum cutoff}
\label{sec:deld4}

So far we have regularized infrared divergences by using dimensional
regularization and have checked
 exponentiation in impact parameter space at the leading and first
 subleading level in $t/s$ and at different orders in the small-$\epsilon$
 expansion.
We have then obtained  the  $D=4$ results by taking, at the end, the
$\epsilon \rightarrow 0$ limit.

In this section, we try to make a more direct connection with the approach
of~\cite{Amati:1990xe} by deriving again $\delta_2$ while staying all
the time in $D=4$ supplemented with a low-momentum cutoff.  We will
show that the $D=4$ result for the real part of $\delta_2$ agrees with
the one obtained in the previous section while this does not appear to
be the case for its imaginary part. We will give an interpretation for
these two contrasting results.

We will start again from the exact expression~\eqref{A1} and first perform a small-$\epsilon$ expansion for a generic kinematics.
A straightforward calculation leads to:
\begin{eqnarray}
{\cal{A}}^{(1)} &=& \frac{1}{ \epsilon} \left[s \log \frac{-s}{\mu^2}   + t \log \frac{-t}{\mu^2} + u \log \frac{-u}{\mu^2}    \right] \nonumber \\
&+&  \left[  u \log \frac{-s}{\mu^2}  \log \frac{-t}{\mu^2}  + 
t \log \frac{-s}{\mu^2}  \log \frac{-u}{\mu^2} + s \log \frac{-u}{\mu^2}  \log \frac{-t}{\mu^2}    \right]\;,
\label{EPS2}
\end{eqnarray}
which agrees with the known result (see, e.g. ~\cite{Dunbar:1994bn}).
As in the previous sections we specify 
 the Riemann sheet along the positive real $s$-axis by taking $\log (-s) = \log s - i \pi$. Using also $s+t+u=0$  to eliminate $u$ we get:
\begin{equation}
{\cal{A}}^{(1)}= -i\pi s (\frac{1}{\epsilon} - \log \frac{-t}{\mu^2}) -i \pi t \log \frac{s+t}{-t} - s \log \frac{s}{s+t}(\frac{1}{\epsilon} - \log \frac{-t}{\mu^2})   +t \log \frac{-t}{s+t}
(\frac{1}{\epsilon} - \log \frac{s}{\mu^2})\,.
\label{EPS6}
\end{equation}
Up to now this expression is exact. We now expand it for $s\gg|t|$ keeping only  terms up to ${\cal O}(t)$ (and neglecting those of  ${\cal O}(t^2/s)$) to get
\begin{equation}
{\cal{A}}^{(1)}= -i\pi s (\frac{1}{\epsilon} - \log \frac{-t}{\mu^2}) -i \pi t \log \frac{s}{-t} + t (\frac{1}{\epsilon} - \log \frac{-t}{\mu^2})   +t \log \frac{-t}{s}
(\frac{1}{\epsilon} - \log \frac{s}{\mu^2})\,.
\label{EPS6b}
\end{equation}
 As a double check,  we can extract  the terms of order $\frac{1}{\epsilon}$ and of order $\epsilon^0$ of  Eq. \eqref{SUM4} and  show that Eq. \eqref{EPS6b} is exactly reproduced.
 
We now get rid of $\epsilon$ by introducing an infrared momentum cutoff $\lambda$ through the relation:
\begin{eqnarray}
\frac{1}{\epsilon} \equiv  \log \frac{\lambda^2}{\mu^2}~~ \Rightarrow \frac{1}{\epsilon} - \log \frac{-t}{\mu^2} = -\log \frac{-t}{\lambda^2} ~~;~~
\frac{1}{\epsilon} - \log \frac{s}{\mu^2} = -\log \frac{s}{\lambda^2} \, .
\label{EPS4}
\end{eqnarray} 
We then arrive at
\begin{eqnarray}
{\cal{A}}^{(1)} \sim i\pi (s+t) \log \frac{-t}{\lambda^2} - t \log \frac{-t}{\lambda^2} \left( \log \frac{s}{\lambda^2} -1 \right) - i \pi t \log \frac{s}{\lambda^2} +t \log^2 \frac{s}{\lambda^2}
\label{EPS7}
\end{eqnarray}
and note that all dependence on $\mu$ has also disappeared as a consequence of UV finiteness.

This gives, for the one-loop amplitude,
\begin{equation}
 \frac{iA^{(1)}}{2s}  =  \frac{iA_L^{(0)}}{2s} \frac{G}{\pi\hbar} {\cal{A}}^{(1)}  
\sim  \,-\, \frac{4\pi  G^2 s^2}{q^2} \log \frac{q^2}{\lambda^2} - 4 \pi G^2 s \log \frac{s}{q^2}  
 +~ 4i G^2 s\log \frac{q^2}{\lambda^2} \left( \log \frac{s}{\lambda^2} -1 \right)\,, 
\label{EPS8}
\end{equation}
where we have used (\ref{LE5}) for the tree amplitude $A_L^{(0)}$.

Using formulae from Appendix \ref{AppB} the ($D=4$) Fourier transform  of the first term is given by
\begin{eqnarray}
\int \frac{d^2 q}{(2\pi\hbar)^2} {\rm e}^{iqb/\hbar} \left( \,-\,  \frac{4\pi  G^2 s^2}{q^2} \log \frac{q^2}{\lambda^2} \right) 
= \,-\,  \frac{1}{2} \left(G s\over \hbar \right)^2 \log^2\left(\frac{ b^2 \lambda^2}{\hbar^2}\right) = \frac{1}{2} (2i \delta_0)^2\,,
\label{EPS9}
\end{eqnarray}
where, in this section, $2i \delta_0$ is the $\epsilon \rightarrow 0$ limit of the
Fourier transform of the tree amplitude $ iA^{(0)}_L/(2s)$ given
in  Eq. (\ref{LE5}) with the above identification $\epsilon^{-1} =
\log(\lambda^2/\mu^2)$.
The Fourier transform of the second term in~\eqref{EPS8} gives
\begin{eqnarray}
\int \frac{d^2 q}{(2\pi\hbar)^2} {\rm e}^{iqb/\hbar} \left(-4\pi G^2 s \log \frac{q^2}{s} \right) = Im (2\Delta_1) = \frac{4 G^2 s}{b^2}\;,
\label{EPS10}
\end{eqnarray}
Finally, the Fourier transform of the third term is equal to
\begin{eqnarray}
\int \frac{d^2 q}{(2\pi\hbar)^2} {\rm e}^{iqb/\hbar} \left(4i G^2 s \log \frac{q^2}{\lambda^2} \right)\left( \log \frac{s}{\lambda^2} -1 \right)  =2 i Re \Delta_1 = \,-\, \frac{4 i G^2 s}{\pi b^2} \left( \log \frac{s}{\lambda^2} -1 \right).
\label{EPS11}
\end{eqnarray}
In conclusion, we have checked (to this order) the exponentiation of the leading eikonal and we have determined the real and imaginary part of $\Delta_1$ that we rewrite here:
\begin{eqnarray}
Re (2 \Delta_1) = \,-\, \frac{4G^2 s}{\pi b^2} \left( \log \frac{s}{\lambda^2} -1 \right) ~~;~~
Im (2 \Delta_1) =  \frac{4 G^2 s}{b^2} \,. 
\label{EPS12}
\end{eqnarray}
Comparing the above results with  the $\epsilon \rightarrow 0$ limit
of those obtained in~\eqref{SUM6a} and~\eqref{SUM7} we note that
there is agreement in the latter case ($Im (2 \Delta_1)$) but not in
the former ($Re (2 \Delta_1)$). The mismatch looks quite substantial
since~\eqref{SUM6a} produces, as $\epsilon \rightarrow 0$, a $\log b^2$
term which is clearly absent in~\eqref{EPS12}. We will argue that the
origin of these two contrasting results  is related to the fact that $Im
(2 \Delta_1)$ is infrared finite while $Re (2 \Delta_1)$ is infrared
divergent.

As a first guess one might argue that we have taken too quickly the
$\epsilon \rightarrow 0$ limit in computing the subleading one-loop
amplitude.  This however is not the case: a direct expansion of the
one-loop amplitude~\eqref{SUM4} shows that no $\log^2 q^2$ term is
generated in the $\epsilon \rightarrow 0$ limit.  As a consequence,
a two-dimensional Fourier transform cannot produce a $\log b^2$
contribution. Therefore the reason for the discrepancy must be found
in the order in which one performs the Fourier transform itself. And
indeed, if one performs {\it first} the Fourier transform in $2-2\epsilon$
dimensions and {\it then} takes the limit, the $\log b^2$ term does come
out as in~\eqref{SUM6a}.  The relevant maths is perfectly exemplified
by the function:
\begin{equation}
  \label{example}
 \frac{q^{- 2 \epsilon}}{\epsilon^2} \left( \epsilon \log (q^2/s) + 2 \right)
\end{equation}
whose $\epsilon \rightarrow 0$ limit has  $\log q^2$ but no $\log^2 q^2$
terms, and whose Fourier transform at finite $\epsilon$ develops a $\log
b^2$ contribution in that same limit.

Our conclusion is that for infrared-divergent terms one has to work all
the time within a consistent regularization scheme. One such scheme is
usually assumed to be dimensional regularization --which we have also
adopted-- while introducing a straight momentum cutoff is not obviously
a consistent scheme (one could try instead to work in a finite box and
then take the limit as done in lattice gauge theories). In any case one
should compare physical infrared-finite quantities in both schemes. For
these reasons in the rest of this section we shall limit ourselves to
the calculation of $Re (\delta_2)$ for which only the knowledge of the
infrared-safe $Im (\Delta_1)$ is needed.

Starting from~\eqref{LE3} and  keeping only terms up to order $\epsilon^0$
in ${\cal A}^{(1)}$ and the leading terms in $F^{(\ell)}$ we have at
order $G^3$
\begin{equation}
\frac{iA^{(2)}}{2s} = \frac{iA_L^{(0)}}{2s} \frac{G^2}{\pi^2\hbar^2} \left( \frac{1}{2} ({\cal{A}}^{(1)})^2 + F^{(2)} \right)\;,
\label{EPS13}
\end{equation}
which should be compared with the corresponding expansion of the eikonal exponentiation
\begin{eqnarray}
(1+ 2i \Delta_1){\rm e}^{2i \delta_0  + 2i \delta_2} \sim \frac{1}{3!} (2i \delta_0)^3 - (2\delta_0) (2\Delta_1) + 2 i \delta_2 + \dots
\label{EPS14}
\end{eqnarray}
For the reasons explained above we will compare only the imaginary part of
these two equations.  Starting from the first term in Eq. (\ref{EPS13}),
we have
\begin{equation} \label{EPS15}
  \frac{iA_L^{(0)}}{2s} \frac{G^2}{2\pi^2\hbar^2} ({\cal{A}}^{(1)})^2
     \sim  
\! \,-\, \frac{2 \pi i G^3 s^3}{\hbar q^2} \log^2 \frac{q^2}{\lambda^2}
  \\    +  {4 G^3 s^2 \over \hbar} \left[  i \pi   \log^2 \frac{q^2}{\lambda^2} - i \pi  \log \frac{q^2}{\lambda^2} \log \frac{s}{\lambda^2} \right],
\end{equation}
where we focused on the leading and the first subleading contributions in the Regge limit.

We can extract the expression for the second term in Eq. (\ref{EPS13})
from~\cite{Henn:2019rgj}. By focusing on the imaginary terms that
are the relevant ones at high energy one gets\footnote{The
terms relevant at high energy can be extracted from Eq.~(6.1) of
Ref.~\cite{Henn:2019rgj}, 
where all the factors of $\log(x)$ should be
replaced by $\log(x)-i\pi$.  
Then one can check
that Eq.~(6.1) of Ref.~\cite{Henn:2019rgj} agrees with the result
of~\cite{BoucherVeronneau:2011qv}. We would like to thank J. M. Henn
for a clarifying discussion on this point.}
\begin{equation}
  \frac{i A_L^{(0)}}{2s} \frac{G^2}{\pi^2\hbar^2} F^{(2)}  \sim  
\,-\, { 2 \pi i G^3 s^2 \over \hbar} 
 \left( \log \frac{q^2}{\lambda^2} - \log \frac{s}{\lambda^2} \right)^2 
\,-\, { 4 \pi i G^3 s^2 \over \hbar}  \log q^2 \, .
\label{EPS23}
\end{equation}
The leading term in $s$ comes from the first term in~\eqref{EPS15} whose  
Fourier transform is 
\begin{eqnarray}
\int \frac{d^2 q}{(2\pi\hbar)^2} {\rm e}^{iqb/\hbar} \left( \,-\, \frac{2i \pi G^3 s^3}{\hbar q^2 } \log^2 \frac{q^2}{\lambda^2} \right) 
=\frac{i}{3!} \left( {G s\over \hbar} \log \left(\frac{ b^2 \lambda^2}{\hbar^2}\right)\right)^3 = \frac{1}{3!} (2i\delta_0)^3
\label{EPS16}
\end{eqnarray}
in agreement with the first term of Eq. (\ref{EPS14}). Note that in the subleading terms the contributions proportional to $\log q^2 \log s$ cancel. The Fourier transform of the rest gives
\begin{equation}
  \int \frac{d^2 q}{(2\pi\hbar)^2} {\rm e}^{iqb/\hbar} 
\Big( {2 \pi i G^3 s^2 \over \hbar} \log^2 \frac{q^2}{\lambda^2} 
\,-\, { 4 \pi iG^3 s^2 \over \hbar} \log \frac{q^2}{\lambda^2}  \Big)
     = \frac{4i G^3 s^2}{\hbar b^2} \log\left(\frac{ b^2 \lambda^2}{\hbar^2}\right)
+ \frac{4i G^3 s^2}{\hbar b^2}.
\label{EPS17}
\end{equation}
Using now:
\begin{eqnarray}
(2i \delta_0) (- Im 2\Delta_1) =   \frac{4i G^3 s^2}{\hbar b^2} \log \left(\frac{ b^2 \lambda^2}{\hbar^2}\right)\, ,
\label{EPS19}
\end{eqnarray}
as well as the imaginary part of~\eqref{EPS14},  we immediately find:
\begin{equation}
  Re (2 \delta_2) =  \frac{4 G^3 s^2}{\hbar b^2}\;.
\label{EPS21}
\end{equation}
Happily, the value of $Re (2 \delta_2) $ coincides with the one obtained in (\ref{SUM11}) and with Eq. (5.26) of Ref.~\cite{Amati:1990xe}. 
(As expected, a similar agreement does not hold for the term of order $\epsilon^0$ in $Im (2 \delta_2)$ whose explicit calculation we omit.)

We finally note that, if we expand up to  order $\epsilon^0$  the quantity
\begin{eqnarray}
(2i\delta_0) (- Im (2\Delta_1)) = \,-\, \frac{4i G^3 s^2 (\pi b^2)^{3\epsilon} \Gamma^3 (1-\epsilon) (1+\epsilon)}{\epsilon \hbar b^2}
\label{EPS30}
\end{eqnarray}
needed in the calculation of $Re(2 \delta_2) $, we get
\begin{eqnarray}
- \frac{4i G^3 s^2}{\hbar b^2} \left(\frac{1}{\epsilon} + 3  \log (\pi b^2) - 3 \psi(1) +1 \right)
\label{EPS31}
\end{eqnarray}
whose term with $\log b^2$ differs by a factor $3$ from the one
of (\ref{EPS19}), while, as mentioned, the results for $Re(2
\delta_2) $ agree. This is due to the following reason: the
interference terms $\epsilon \times 1/\epsilon$ that we neglected in
calculating~\eqref{EPS19} in $D=4$ are identical to the corresponding
interference terms neglected in~\eqref{EPS15}. This happens because the
$1/\epsilon$ contribution is a constant in both cases and so the Fourier
transform acts non-trivially only on the ${\cal O}(\epsilon)$ term mapping
exactly the ${\cal O}(\epsilon)$ contribution of ${\cal{A}}^{(1)}$ into
that of $\Delta_1$. Once more the same cancellation does not occur for
$Im (2 \delta_2)$.

\section{Summary and Outlook}
\label{sec:concl}

Four-point amplitudes in ${\cal N}=8$ supergravity are known with a great
degree of precision. In this work we set up a systematic approach for the
analysis of these loop amplitudes in the Regge regime where the momentum
transferred is much smaller than the centre of mass energy. A first result
is that, even in this highly supersymmetric setup, some of the contributions
that grow polynomially with the energy are not accounted for by the
exponentiation of the leading eikonal~\eqref{LE6} alone. Instead they give rise
to a new classical contribution ($2i \delta_2$ in~\eqref{eq:dDsum})
that modifies the leading eikonal at 3PM order, {\rm i.e.} $(R/b)^2$
in $D\to 4$ and in the Regge regime $R\ll b$, where $b$ is the
impact parameter and $R$ is a scale related to the energy of the
process~\eqref{defR}. Corrections at 2PM order are absent in massless
theories, see the comment after~\eqref{scalinggenl}, but it is interesting
to notice that in a maximally supersymmetric setup, they are absent also
when the external states are massive~\cite{Caron-Huot:2018ape}.
Our results show that this
cancellation, motivated by supersymmetry, does not survive at higher
orders when both particles are dynamical. Further corrections at 5PM
order, {\rm i.e.} $(R/b)^4$, are expected and should be extracted from
the sub-subleading terms in the four-loop amplitude.

Notice that these power-like
contributions are different from the most logarithmically enhanced
terms discussed in~\cite{Bartels:2012ra,SabioVera:2019edr}. In
theories with only spin $1$ particles, the dominant terms in the Regge
limit are proportional to $(\log^2(t/s))^\ell$ at $\ell$ loops. By
contrast, in gravity theories these terms take the form
$(t\log^2(t/s))^\ell$, and thus become increasingly power-suppressed
in $t$ as the loop order increases. Nevertheless, an algorithm exists
for deriving them at arbitrary
order~\cite{Bartels:2012ra,SabioVera:2019edr}, and they should be
resummed in order to describe the scattering process for values of the
impact parameter $b$ that are closer to $R$ (even before reaching
Planckian scales).

The main property of the classical eikonal is that it should
exponentiate, see in~\eqref{intro1}: in this way the full amplitude has
the expected classical limit $\hbar\to 0$, where the only singular term
is a WKB-like exponential, see \cite{Kosower:2018adc} for a closely
related discussion. Contrary to what happens for the leading eikonal
and for the 2PM correction when this is present, the 3PM result
$(2\delta_2)$ contains both a real and an imaginary part. The real
part is directly related to physical observables such as the deflection
angle and the Shapiro time delay and so one would expect it to be free
of IR divergences. This is the case in our result since the infrared
divergent term in the real part of the two-loop amplitude is cancelled
in the subtraction~\eqref{eq:d2deta} yielding an IR finite result for
$Re(2\delta_2)$. The imaginary part of the eikonal is IR divergent and
it would be very interesting to study a physical observable, such as
an inclusive cross section, which is sensitive to $Im(2\delta_2)$, so
as to check how the cancellation of the IR divergences works at higher
order, generalising for instance the discussion after Eq.~\eqref{vsuppr}
at two loops.

There is another interesting aspect related to IR divergences that we
analysed in some detail: the relation between the IR exponentation in
momentum space~\eqref{LE3} and the eikonal exponentiation in impact
parameter space~\eqref{intro1}. At leading level in the Regge regime
the two expressions match in a non-trivial way in the common regime of
validity for any value of the dimensional regularisation parameter
$\epsilon$ as already discussed~\cite{DiVecchia:2019myk}. The leading
eikonal is universal, {\rm i.e.} it does not depend on the presence of
supersymmetry and is the same for all gravity theories that at large
distances reduce to CGR. Then the relation between the two
exponentiations provides an easy set of predictions for the terms of
the $\ell$-loop gravitational amplitudes that scale as $s^{\ell+1}$
for small $t$. In this paper we extended this logic to the subleading
terms in the Regge regime. At this order the amplitudes depend on the
details of the theory and we focused on the case of ${\cal N}=8$
supergravity.\footnote{However it is interesting to notice that
  $Re(\delta_2)$ seems to be universal, see the comment
  after~\eqref{SUM11}.} By using the explicit results
of~\cite{Henn:2019rgj} we compared the two exponentiations at all
loops for the first four terms in the $\epsilon \to 0$ expansion. As
discussed in section~\ref{sec:allorders}, there is an impressive
agreement between the eikonal prediction~\eqref{intro1} and the
explicit results of~\cite{Henn:2019rgj} that satisfy perfectly the IR
exponentation in momentum space~\eqref{LE3}. However there is a
mismatch for one term appearing at the lowest power of $1/\epsilon$
and the lowest power of $\log(q^2)$ accessible with the current
data. At three-loop order the mismatch appears in the IR finite part,
see~\eqref{eq:mismatch}: then a correction in the ${\cal O}(\epsilon)$
part of the 3PM eikonal or the finite part of the three-loop amplitude
can restore the agreement at three loops between the two
exponentiations. However the tension resurfaces at four loops and
higher in the terms ${\cal O}(\epsilon^{4-\ell})$,
see~\eqref{eq:mismatch2}. What is most puzzling is that such a
mismatch indicates a breakdown of either the eikonal or the IR
exponentation. It may be that one has to restrict the comparison of
the two results only to physical/IR finite observables. Understanding
this point better is of course of great interest and would probably
require to specify better the regime of validity of both formulas. The
standard approach to amplitude calculations is to fix the kinematics,
including the Mandelstam variables, and take the small $\epsilon$
expansion to focus around $D=4$. This implies that the IR regulator is
the smallest scale in the problem. In the eikonal approach we kept
$\epsilon$ fixed (even when small) and then considered all values of
the exchanged momentum $|q|$: actually the most important
contributions to the large distance physics ($b\gg R$) relevant to the
Regge regime are those that are divergent as $|q|\to 0$.  It would be
interesting to understand whether the discrepancy mentioned above is
related to the different kinematics where the two exponentiations are
valid~\footnote{See
  ref.~\cite{Bret:2011xm,DelDuca:2011ae,Caron-Huot:2017zfo,Caron-Huot:2017fxr}
  for studies relating the exponentiation of infrared singularities to
  known properties of the Regge limit in a gauge theory
  context.}. Clarifying this point may be relevant beyond the ${\cal
  N}=8$ case studied in this work, since now, even for the physically
interesting case of the massive scattering in CGR, the focus is on 3PM
and higher order corrections~\cite{Bern:2019nnu, Bern:2019crd} where
such subtleties may play some role.

\vspace{5mm}

\noindent {\large \textbf{Acknowledgements} }

We thank Emil Bjerrum-Bohr, Marcello Ciafaloni, Arnau Koemans Collado, Andrea Cristofoli, Poul Henrik Damgaard, Einan Gardi, Henrik Johansson, Marios Hadjiantonis, Carlo Heissenberg, Johannes Henn and Andr\'{e}s
Luna for useful conversations.  The research of SGN is supported in
part by the National Science Foundation under Grant No.~PHY17-20202.
CDW and RR are supported by the UK Science and Technology Facilities
Council (STFC) Consolidated Grant ST/P000754/1 ``String theory, gauge
theory and duality, and/or by the European Union Horizon 2020 research
and innovation programme under the Marie Sk\l{}odowska-Curie grant
agreement No. 764850 ``SAGEX''.  PDV, RR and GV would like to thank
the Galileo Galilei Institute for hospitality during the workshop
``String theory from the worldsheet perspective'' where they started
discussing this topic.  PDV was supported as a Simons GGI scientist
from the Simons Foundation grant 4036 341344 AL. The research of PDV
is partially supported by the Knut and Alice Wallenberg Foundation
under grant KAW 2018.0116.

\appendix

\section{Useful Fourier transforms to impact parameter space}
\label{AppB}

In this Appendix we derive the Fourier transforms into impact parameter
space that we have used in this paper. The basic starting formula
is~\footnote{For non integer values of $D-2$ this is defined, as usual,
via analytic continuation from all positive integer values of that
same quantity.}:
\begin{equation}
  \int \frac{d^{D-2} q}{(2\pi\hbar)^{D-2}} {\rm e}^{ibq/\hbar} 
\left(q^2\over \hbar^2\right)^\nu  
= \frac{2^{2\nu}}{\pi^{1-\epsilon}} \frac{\Gamma (1+\nu -\epsilon)}{\Gamma (- \nu) (b^2)^{\nu +1 - \epsilon}}~~;~~ D-4 =-2\epsilon 
  \label{B1} \, .
\end{equation}
It can be rewritten as follows:
\begin{eqnarray}
\int \frac{d^{D-2} q}{(2\pi\hbar)^{D-2}} {\rm e}^{ibq/\hbar} \sum_{n=0}^{\infty} \frac{\nu^{n}}{(n+1)!}  \log^{n+1} 
\left(q^2\over \hbar^2\right) =
- \frac{f(\nu) (\pi b^2)^\epsilon}{\pi b^2}  \sum_{n=0}^{\infty} \frac{ (-1)^n  \nu^n}{n!} \log^n b^2\,,
\label{B2}
\end{eqnarray}
where
\begin{eqnarray}
f (\nu ) = 2^{2\nu} \frac{\Gamma (1+\nu)}{\Gamma (1- \nu)} \equiv  \sum_{m=0}^\infty \frac{f^{(m)}}{m!} \nu^m.
\label{B3}
\end{eqnarray}
The first few coefficients of the above sum are:
\begin{eqnarray}
f^{(0)} =1~~;~~f^{(1)} = \log 4 + 2 \psi(1)~~;~~f^{(2)} =( \log 4 + 2 \psi(1))^2 ~~;~~ \psi(1) = - \gamma_E.
\label{B4}
\end{eqnarray}
Inserting the expansion in Eq. (\ref{B3}) in Eq. (\ref{B2}) we get
\begin{eqnarray}
\int \frac{d^{D-2} q}{(2\pi\hbar)^2} {\rm e}^{ibq/\hbar}  
\log^{n+1} 
\left(q^2\over \hbar^2\right) 
=- (n+1) \frac{(\pi b^2)^\epsilon}{\pi b^2}  \sum_{m=0}^n \left( \begin{array}{c} n \\ m \end{array} \right) f^{(m)} (-1)^{n-m} \log^{n-m} b^2.
\label{B5}
\end{eqnarray}
For $n=0$ we get
\begin{eqnarray}
\int \frac{d^{D-2} q}{(2\pi\hbar)^{D-2}} {\rm e}^{ibq/\hbar}  \log  
\left(q^2\over \hbar^2\right) 
= - \frac{(\pi b^2)^{\epsilon}}{\pi b^2}~~,\qquad\qquad ~~ f^{(0)} =1 \,.
\label{B6}
\end{eqnarray}
For $n=1$ we get
\begin{eqnarray}
\int \frac{d^{D-2} q}{(2\pi\hbar)^{D-2}} {\rm e}^{ibq/\hbar}  \log^2  
\left(q^2\over \hbar^2\right)
&=& - \frac{2(\pi b^2)^\epsilon}{\pi b^2}  
  \left(- f^{(0)}  \log b^2 + f^{(1)}   \right)= \frac{2 (\pi b^2)^\epsilon}{\pi b^2} \log \frac{b^2}{e^{f^{(1)}}}
\nonumber \\
&=&\frac{2 (\pi b^2)^\epsilon}{\pi b^2}\left( \log \frac{b^2}{4} - 2 \psi (1) \right).
\label{B7}
\end{eqnarray}
For $n=2$ we get
\begin{eqnarray}
\int \frac{d^{D-2} q}{(2\pi\hbar)^2} {\rm e}^{ibq/\hbar}  \log^3  
\left(q^2\over \hbar^2\right)
&=& -  \frac{3(\pi b^2)^\epsilon}{\pi b^2}  \left( f^{(0)}  \log^2 b^2 - 2 f^{(1)}  \log b^2 + f^{(2)} \right) \nonumber \\
&=& -  \frac{3(\pi b^2)^\epsilon}{\pi b^2}\left( \log^2 \frac{b^2}{e^{f^{(1)}}}  - (f^{(1)})^2 + f^{(2)}\right)
\nonumber \\
&=& -  \frac{3(\pi b^2)^\epsilon}{\pi b^2}  \log^2 \frac{b^2}{4 {\rm e}^{2 \psi (1)}}. 
\label{B8}
\end{eqnarray}
In the main text we are also using the inverse Fourier transform (from $b$
to $q$-space) which can be easily derived from the above results using
the well known properties of the Fourier transform.  As an example the
analog of (B.1) reads:
\begin{equation}
  \int\! { d^{D-2} b} \, {\rm e}^{-ibq/\hbar} (b^2)^{-\nu} 
= \frac{\pi^{\frac{D-2}{2}}}{2^{2 \nu + 2 -D}} \frac{\Gamma\left(\frac{D}{2}-1-\nu\right)}{\Gamma(\nu)}
\left(q^2\over \hbar^2\right)^{1+\nu-\frac{D}{2}}
  \label{B1bis}
\end{equation}
from which we can derive another useful relation 
\begin{align} 
  \int\! { d^{D-2} b} \, {\rm e}^{-ibq/\hbar} (b^2)^{-\nu} \log b^2 
&= \frac{\pi^{\frac{D-2}{2}}}{2^{2 \nu + 2 -D}} \frac{\Gamma\left(\frac{D}{2}-1-\nu\right)}{\Gamma(\nu)}
\left(q^2\over \hbar^2\right)^{1+\nu-\frac{D}{2}}
\nonumber\\
&\qquad\qquad\times \left[ \log\left( \frac{4\hbar^2}{q^2} \right) + \psi( \textstyle{\frac{D}{2}} -1-\nu) + \psi(\nu)
\right]\,.
  \label{logb}
\end{align}

\section{Results of Henn and Mistlberger
}
\label{AppC}

In this appendix we write the eikonal limit of the three-loop ${\cal N}=8$
4-point amplitude recently derived in~\cite{Henn:2019rgj} up to order
$\epsilon^0$ in dimensional regularization. We write also the two-loop
result up to order $\epsilon^2$ included in the same paper. With  respect
to~\cite{Henn:2019rgj}, we write the result in the $s$-forward channel,
{\rm i.e.} with $s>0$ and $t,u<0$ and, for simplicity, in the equations
below we set the dimensional regularization scale to one $\mu=1$. As
mentioned in the main text, we focus only on the non-analytic terms
as $|t|=q^2 \to 0$ as they are the only ones yielding a long-range
interaction in the impact parameter space and so are captured by the
eikonal exponentiation~\eqref{intro1}.  We organise the formulas by
writing first the leading eikonal terms (proportional to $s^\ell$ at
$\ell$ loops) and then the first subleading term. For each of the two
contributions we order the various terms according to the power $n$
of $\log^n(q^2)$.

At two-loop order we have 
\begin{align}
 {\cal A}^{(2)} = & ~ s^2 \left\{- \frac{\pi^2}{3} \epsilon^2
\log^4(q^2)+\frac{2 \pi^2}{3}\epsilon  \log^3(q^2) - \pi^2 \log^2(q^2)
+ \log(q^2) \left(\frac{\pi^2}{\epsilon }-6 \pi^2 \zeta_3
\epsilon^2\right) \right\}
 \nonumber \\   \label{eq:a2L}
 + & ~  s q^2 \left\{\frac{2 i \pi}{15} \epsilon^2
\log^5(q^2)+\log^4(q^2) \left[\left(\frac{\pi^2 }{3}+\frac{4 i \pi
}{3}\right) \epsilon^2-\frac{1}{3} i \pi   \epsilon \right]\right.
 \\ \nonumber &
  +\log^3(q^2) \left[\epsilon^2 \left(-\frac{7 i \pi^3}{9} -\frac{4
\pi^2 }{3}-\frac{8 i \pi  }{3}-\frac{8 i \pi}{3} \log
s\right)+\left(-\frac{ 2 \pi^2}{3} -\frac{4 i \pi  }{3}\right)
\epsilon +\frac{2 i \pi  }{3}\right]
 \\ \nonumber &
  +  \log^2(q^2) \left[\epsilon^2 \left(41 i \pi   \zeta_3+5 i \pi^3
\right)-\frac{i \pi  }{\epsilon }+\epsilon  \left(\frac{7 i \pi^3}{6}
+2 \pi^2 +4 i \pi  +4 i \pi   \log s\right)+\pi^2 \right]
 \\ \nonumber &
  + \log(q^2) \left[+\frac{i \pi  }{\epsilon^2}+\frac{-\pi^2 +2 i \pi
}{\epsilon }-\frac{7 i \pi^3}{6}  -2 \pi^2 -4 i \pi  -4 i \pi   \log
s
 \right . \\ \nonumber & \left. \left.
   \quad - \epsilon  \left(35 i \pi   \zeta_3+5 i \pi^3 \right) +
\epsilon^2 \left(-86 i \pi   \zeta_3-12 i \pi   \zeta_3 \log
s-\frac{31 i \pi^5}{15}  \right)\right] \right\}+\ldots
\end{align}
and at three-loop order we have 
\begin{align}
  \label{eq:a3L}
 {\cal A}^{(3)} = & ~ s^3 \left\{-\frac{3 i \pi^3}{4}
\log^3(q^2)+\frac{3 i \pi^3 \log^2(q^2)}{4 \epsilon } - \frac{i \pi^3
\log(q^2)}{2 \epsilon^2}\right\}
\\ \nonumber + & ~ s^2 q^2 \left\{-\frac{9\pi^2}{8} \log^4(q^2)
+\log^3(q^2) \left[\frac{5 \pi^2}{4 \epsilon }+\frac{3 i \pi^3}{4}
-\frac{3 \pi^2}{4}+\frac{3}{4} \pi^2 \log s \right]\right.
 \\ \nonumber &
+ \log^2(q^2) \left[-\frac{\pi^2 }{\epsilon^2}-\frac{3}{4} \frac{i
\pi^3 + \pi^2+ \pi^2  \log s}{\epsilon }+\frac{11 \pi ^4 }{8}-\frac{9
i \pi^3}{4} +\frac{9 \pi^2 }{2}+\frac{9\pi^2}{2} \log s \right]
 \\ \nonumber &
  + \log(q^2) \left[\,\frac{\pi^2 }{2 \epsilon^3}+\frac{i \pi^3 +3
\pi^2 + \pi^2   \log s}{2 \epsilon^2}
\right.  \\ \nonumber & \left.\left.
    \qquad  \qquad     +\frac{\frac{3 i \pi^3}{2} -3 \pi^2 -\frac{11
\pi ^4 }{12}-3 \pi^2  \log s}{\epsilon }-\frac{35 \pi^4 }{6}-\frac{91
\pi^2}{2}  \zeta_3\right]\right\} +\ldots~.
\end{align}
Finally, we can derive the IR divergent part of the four-loop amplitude
by using the exponentiation~\eqref{LE3}. For this it is sufficient to
know the two-loop remainder function up to order $\epsilon$ and that of
three-loop remainder function at order ${\cal O}(\epsilon^0)$ and both
these results are provided in the ancillary files of~\cite{Henn:2019rgj}.
Once translated in our $s$-channel convention ($s>0$, $t,u<0$), they read
\begin{align}
\nonumber 
  {F}^{(2)} = & ~ \pi^2 s^2 3 \epsilon \zeta_3 +  \pi s q^2 \left\{-\frac{5 i}{12} \epsilon \log^4(q^2) + \left[\frac{i}{3} +  \epsilon \left(i \log s+ \frac{\pi}{2} - i\right) \right] \log^3(q^2)  \right. \\ \nonumber & + \left[ -\frac{\pi}{2} + i - i \log s + \epsilon \left(-\frac{i \log^2s}{2} + \log s\left(i-\frac{\pi}{2}\right) +\frac{5 i \pi^2}{12}+i+\frac{\pi}{2} \right) \right] \log^2(q^2) \\ \nonumber & \left. + \left[i \log^2s+( \pi\!-\! 2i ) \log s-\! \frac{2i \pi^2}{3}\! -\! \pi\! -\! 2i\! +\! \epsilon \left(-\frac{i \log^3s}{3} + \left(i\! -\! \frac{\pi}{2}\right) \log^2s \right. \right.\right. \\ \label{eq:F2e0} & ~ \left.\left.\left. +  \left(\frac{i \pi^2}{2}\! +\! 2 i \!+\! \pi \right)\log s - 33 i \zeta (3)\! +\! \frac{\pi^3}{6}\! + \! 2 i\! -\! \frac{7 i \pi^2}{2}\! + \! \pi\right)\right] \log(q^2)\right.\\ \nonumber & ~ +
\frac{1}{6} \log s \left((-2 i \log s-3 \pi +6 i) \log s +4 i \pi^2+6 \pi +12 i \right) + 4 i \zeta(3)+\frac{\pi^3}{4}  \\ \nonumber & ~~ +\frac{2 i \pi ^2}{3}+\pi +2 i + \epsilon \left(\frac{\zeta_3}{2} (50 i  \log s +19 \pi +50 i) +\frac{1}{12}  \log s \Big( \log s \Big(3 i \log^2s \right.  \\ \nonumber & ~~ +6 (\pi -2 i) \log s -36 i+\pi  (-18-11 i \pi )\Big)-2 \pi  (18+\pi  (4 \pi -13 i))-72 i\Big)  \\ \nonumber & ~\; \left. \left. +\frac{151 i \pi ^4}{180}+\frac{5 \pi ^3}{6}+\frac{13 i \pi ^2}{6}-3 \pi -6 i\right)  \right\} + \ldots
\end{align}
and
\begin{align}
\nonumber 
  {F}^{(3)} = & ~ - \frac{2}{3} i \pi ^3 s^3 \zeta (3)+ \pi^2 s^2 q^2 \left\{ \frac{1}{12} \log^4(q^2) + \left(-\frac{1}{3} \log s+\frac{i \pi }{6}+\frac{1}{3}\right) \log^3(q^2) \right. \\ \nonumber &  + \left(\frac{\log^2 s}{2}- \left(1+\frac{i \pi}{2}\right) \log s -\frac{\pi ^2}{3}+\frac{i \pi }{2}-1\right) \log^2(q^2)  \\ \label{eq:F3e0} & + \left(-\frac{\log^3s}{3} + \left(1+\frac{ i \pi}{2}\right) \log^2s + \left(2 +\frac{2\pi^2}{3}  -i \pi\right) \log s -4 \zeta (3)\right)\log(q^2)  \\ \nonumber & +\frac{ \log^4 s }{12}-\frac{1}{6} i (\pi -2 i)  \log^3 s -\frac{1}{6} \left(6-3 i \pi +2 \pi ^2\right) \log^2 s -i \pi  (\zeta (3)-1)  \\ \nonumber & \left. ~  +8 \zeta_3 + \frac{5 \zeta_5-2 \zeta_3^2}{\pi ^2}+\frac{2 \pi ^2}{3}-\frac{i \pi ^3}{4}-\frac{841 \pi ^4}{5670}-2 \right\} + \ldots~.
\end{align}
Then the non-analytic part of ${\cal A}_4$ reads
\begin{align}
\nonumber
  {\cal A}^{(4)} = & ~ s^4 \pi^4 \left\{-\frac{\log q^2}{6 \epsilon^3}+\frac{\log^2q^2}{3 \epsilon^2}-\frac{4 \log^3q^2}{9 \epsilon} \right\} \\  + &  ~s^3 q^2 \pi^3 \left\{\frac{10 i}{9 \epsilon}  \log^4q^2 + \left[-\frac{8 i }{9 \epsilon^2}+\frac{4 (\pi -2 i \log s)}{9 \epsilon}\right] \log^3q^2 \right.   \label{eq:a4L} \\ \nonumber & + \left[\frac{i }{2 \epsilon^3}-\frac{\pi \!-\! 2 i \!- \! 2 i \log s}{3 \epsilon^2}-\frac{16 i \log s + 5 i \pi^2 + 8 \pi + 16 i}{6 \epsilon}\right]  \log^2 q^2 \\ \nonumber & +\left. \left[\frac{ -i }{6 \epsilon^4}+\frac{\pi \! - \! 4 i \!-\! 2 i \log s}{6 \epsilon^3}+\frac{16 i \log s \!+ \! 5 i \pi^2 \!+ \! 8 \pi \!+\! 16 i}{12 \epsilon^2}+\frac{173 i \zeta (3)\! +\! 21 i \pi ^2}{6 \epsilon}\right] \log q^2 \right\}.
\end{align}


\providecommand{\href}[2]{#2}\begingroup\raggedright\endgroup

\end{document}